\documentclass[aps, pre, twocolumn, superscriptaddress, showpacs]{revtex4-1}

\usepackage[english]{babel}
\usepackage{hyperref}

\usepackage[pdftex]{graphicx}
\usepackage{color}
\usepackage[toc,page]{appendix}

\usepackage{amsmath}
\usepackage{amssymb}
\usepackage{mathtools}
\usepackage{braket}

\usepackage[sort&compress]{natbib}

\newcommand{\fig}[1]{Figure~\ref{fig:#1}}
\newcommand{\Fig}[1]{Figure~\ref{fig:#1}}

\newcommand{\eq}[1]{Eq.~(\ref{eq:#1})}
\newcommand{\lr}[1]{\ensuremath{\left( #1 \right)}}
\renewcommand{\d}{\mathrm{d}}
\renewcommand{\Re}[1]{\ensuremath{\mathrm{Re} \left(#1\right)}}
\renewcommand{\Im}[1]{\ensuremath{\mathrm{Im} \left(#1\right)}}
\newcommand{\I}{\mathrm{i}}

\newcommand{\mc}{\mathcal}
\newcommand{\Abs}[1]{\ensuremath{\left| #1 \right|}}

\newcommand{\eps}{\epsilon}

\begin{document}

\title{Efficient quantum transport in disordered interacting many-body networks}

\author{Adrian Ortega}
\email{adrianortega@fis.unam.mx}
\affiliation{Instituto de Ciencias F\'isicas, Universidad Nacional Aut\'onoma de M\'exico, Cuernavaca, M\'exico}

\author{Thomas Stegmann}
\email{stegmann@icf.unam.mx}
\affiliation{Instituto de Ciencias F\'isicas, Universidad Nacional Aut\'onoma de M\'exico, Cuernavaca, M\'exico}

\author{Luis Benet}
\email{benet@fis.unam.mx}
\affiliation{Instituto de Ciencias F\'isicas, Universidad Nacional Aut\'onoma de M\'exico, Cuernavaca, M\'exico}
\affiliation{Centro Internacional de Ciencias, Cuernavaca, M\'exico}

\date{03 October 2016}

\pacs{05.60.Gg, 73.23.-b, 87.15.-v}


\begin{abstract}
  The coherent transport of $n$ fermions in disordered networks of $l$ single-particle states
  connected by $k$-body interactions is studied. These networks are modeled by embedded Gaussian
  random matrix ensemble (EGE). The conductance bandwidth as well as the ensemble-averaged total
  current attain their maximal values if the system is highly filled $n \sim l-1$ and $k\sim n/2$.
  For the cases $k=1$ and $k=n$ the bandwidth is minimal. We show that for all parameters the
  transport is enhanced significantly whenever centrosymmetric ensemble (csEGE) are considered. In
  this case the transmission shows numerous resonances of perfect transport. Analyzing the
  transmission by spectral decomposition, we find that centrosymmetry induces strong correlations
  and enhances the extrema of the distributions. This suppresses destructive interference effects in
  the system and thus, causes backscattering-free transmission resonances which enhance the overall
  transport. The distribution of the total current for the csEGE has a very large dominating peak
  for $n=l-1$, close to the highest observed currents.
\end{abstract}

\maketitle

\section{Introduction and motivation}
\label{sec:Introduction}

The motivation of our study is to understand properties of certain biomolecules such as the
Fenna-Mathews-Oslon (FMO) light-harvesting complex
\cite{fennamatthews_nature}, as well as the engineering of quantum devices, like quantum buses based
on spin chains \cite{bose_review}. These systems display a remarkably efficient transport of
electronic excitations or quantum states. In the context of biomolecules for which quantum effects
may play a role, it seems realistic that models describing excitonic transport have to display
disorder \cite{tor_pre, 2014NJP-ZMWB, 2013PRL-walschaers, 2015PRE-walschaers}. Therefore one way to
look at the problem is modeling the system by means of purely stochastic approaches, such as those
of random matrix theory (RMT) \cite{brodyetal, guhr}. Quantum spin chains \cite{qnet_book} are
candidates for playing the role of quantum buses between quantum processors in a quantum
computer. The transfer efficiency of states between the ends of the chain can happen with
probability 1, for a very particular interaction \cite{Plenio2004, christandl_prl, christandl_pra,
  2009IJQI-BCMS}. There are models for treating the random coupling case such as \cite{2005NJP-BB,
  casaccino}, but they do not address the effects of many-body interactions, which we consider using
the embedded random-matrix ensemble \cite{2003JPA-BW,kotabook}.

In our previous work \cite{2015ADP-OrtegaMananBenet} we have studied the efficiency of small bosonic
and fermionic disordered networks generated by the embedded Gaussian random matrix ensemble (EGE)
for closed systems. In this work, we open (in the scattering sense) the fermionic many-body
interacting system and analyze the transmission and current averaged over the ensemble. The question
which we address is how the coherent transport between two states of a small disordered interacting
quantum system can be improved. We concentrate in small networks since the Hilbert space dimension
of FMO molecules and spin chains is rather small \cite{tor_pre, 2014NJP-ZMWB, 2013PRL-walschaers,
  christandl_pra, christandl_prl, 2009IJQI-BCMS, 2015PRE-walschaers}. We will show that the presence
of centrosymmetry \cite{cantoni} in the ensemble is again a fundamental building block to enhance
transport in disordered quantum networks \cite{Plenio2004, christandl_prl, 2013PRL-walschaers,
  2015PRE-walschaers}.

We consider a system of $l$ single-particle states, which are coupled via $k$-body interactions and
occupied by $n$ fermions. The Hamiltonian of this system is taken from embedded Gaussian ensemble of
random matrices (EGE). The transport between two states of the Hilbert space of this system is
calculated by means of the nonequilibrium Green's function (NEGF) method. Details on the model and
our method are provided in Section~\ref{sec:ModelMethods}.

\begin{figure}[t]
  \centering
  \hspace*{-6mm}
  \includegraphics[scale=0.38]{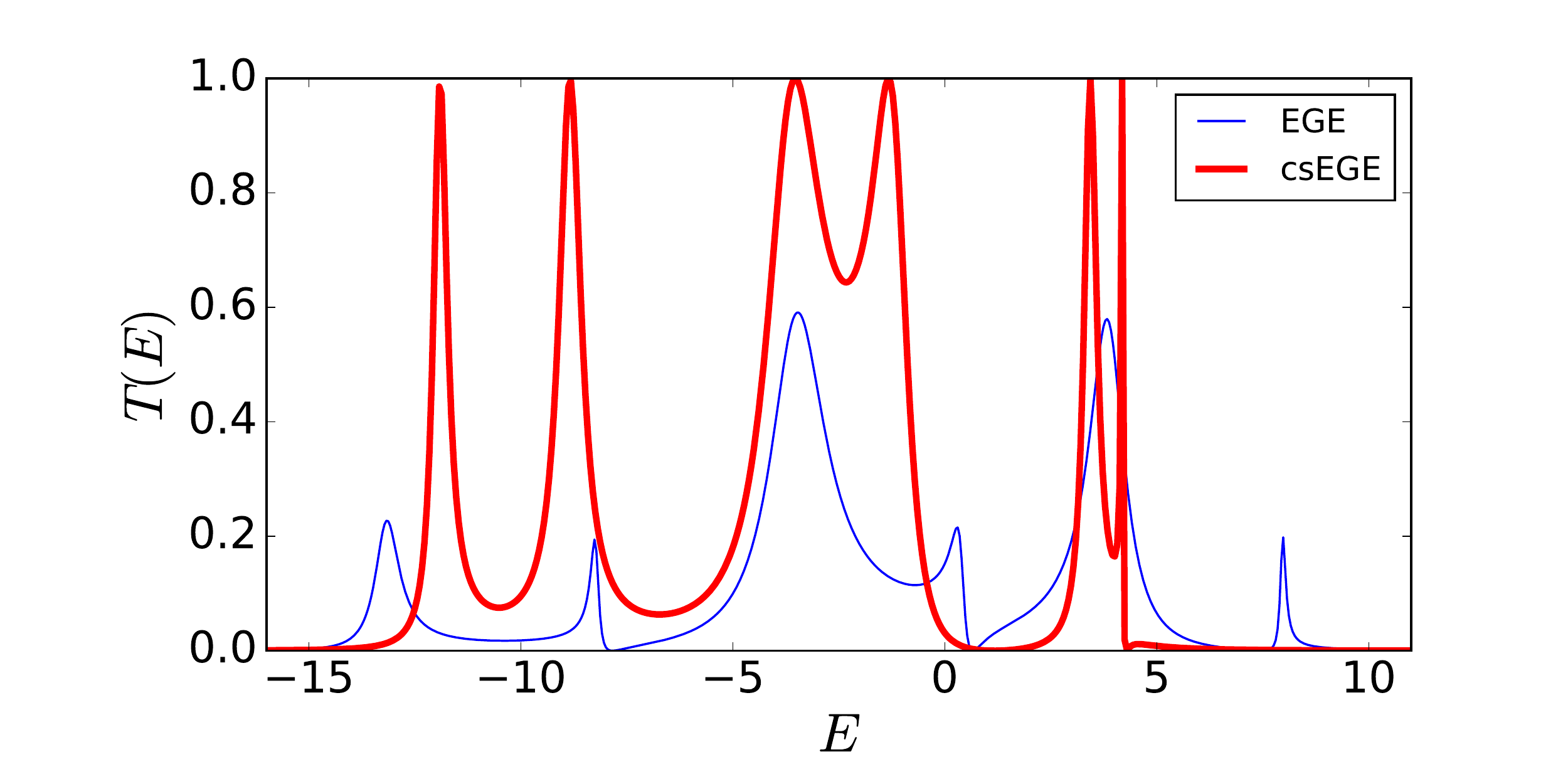}
  \caption{(color online) Transmission $T(E)$ as a function of the energy $E$ for two typical
    members of the random matrix ensemble. The parameters are $l=6$, $n=5$, $k=3$. In both cases
    resonances can be observed which are approximately at the eigenenergies of the Hamiltonian. If
    the Hamiltonian is centrosymmetric (red-thick curve), we observe numerous resonances of perfect
    transport ($T=1$). The transmission increases significantly compared to a Hamiltonian without
    this symmetry property (blue-thin curve).}
  \label{fig:1}
\end{figure}

The energy resolved transmission $T(E)$ for two typical members of the ensemble is shown in
\fig{1}. In both cases the transmission shows resonances which are located approximately at the
eigenenergies of the Hamiltonian. Many resonances of perfect transport ($T=1$) can be observed if
the Hamiltonian is centrosymmetric; see the red curve in \fig{1}. Moreover, our objective here is
not restricted to certain energies but on improving the transport in the whole conduction band,
which is achieved by centrosymmetry.

In Section~\ref{sec:ResultsDiscussion}, we shall discuss in detail the enhancement of the transport
properties due to centrosymmetry as well as its effect on the statistical properties of the
system. In particular, we find that important spectral correlations appear for the open
centrosymmetric Hamiltonian. These, combined with the structure of the eigenfunctions, yield higher
and broader transmission resonances, as well as a wider conduction band, which result in an enhanced
current. The conclusions and outlook are presented in Section~\ref{sec:ConclusionsOutlook}.

\section{Model and methods}
\label{sec:ModelMethods}

\subsection{Embedded random matrix ensemble for disordered interacting systems}
\label{sec:EGE}

We consider a set of $l$ degenerate (fermionic) single-particle states $\ket{j}$, with
$j=1,2,\dots, l$. The associated creation and annihilation operators are $a^\dagger_j$ and $a_j$,
with $j=1,\dots, l$. They obey the usual anti-commutation relations which characterize fermions. We
define the operators that create a normalized state with $k < l$ fermions from the vacuum state as
$\psi^\dagger_{k;\alpha} = \psi^\dagger_{j_1,\dots, j_k} = \prod_{s=1}^k a^\dagger_{j_s}$, with the
convention that the indices are ordered increasingly $j_1<j_2<\cdots < j_k$.  The index $\alpha$ in
the many-body operators labels the specific occupation given by the $j_s$, and simplifies the
notation. The corresponding annihilation operator $\psi_{k;\alpha}$ is constructed analogously.

The random $k$-body Hamiltonian reads
\begin{equation}
  \label{eqHam}
  H_k = \sum_{\alpha, \gamma} v_{k; \alpha, \gamma}
  \psi^\dagger_{k;\alpha} \psi_{k;\gamma}\ ,
\end{equation}
which takes into account interactions between up to $k$ particles. The coefficients
$v_{k; \alpha, \gamma}$ are randomly distributed independent Gaussian variables with zero mean and
unit variance
\begin{equation}
  \label{variance}
  \overline{v_{k; \alpha, \gamma} v_{k; \alpha', \gamma'}} =
  \delta_{\alpha,\gamma'}\delta_{\alpha',\gamma} +
  \delta_{\alpha,\alpha'}\delta_{\gamma,\gamma'}.
\end{equation}
The Hamiltonian $H_k$ acts on a Hilbert space spanned by distributing $n\geq k$ particles on the
$l>n$ single-particle states. A complete set of basis states is given by the set
$\psi^\dagger_{n;\alpha} \ket{0}$. The dimension of the Hilbert space is $N=\binom{l}{n}$. This
defines the $k$-body embedded Gaussian orthogonal ensemble of random matrices for
fermions~\cite{2003JPA-BW, kotabook}.

By construction, the case $k=n$ is identical to the canonical ensemble of random matrix
theory~\cite{guhr}, i.e. to the Gaussian orthogonal ensemble (GOE). For $k < n$, the matrix elements
of $H_k$ may be identical to zero and display correlations. Zeros appear whenever no $k$-body
operator exist that links together the $n$-body states. Correlations arise because matrix elements
of $H_k$ not related by symmetry may be identical.

\begin{figure}[t]
  \includegraphics[scale=0.19]{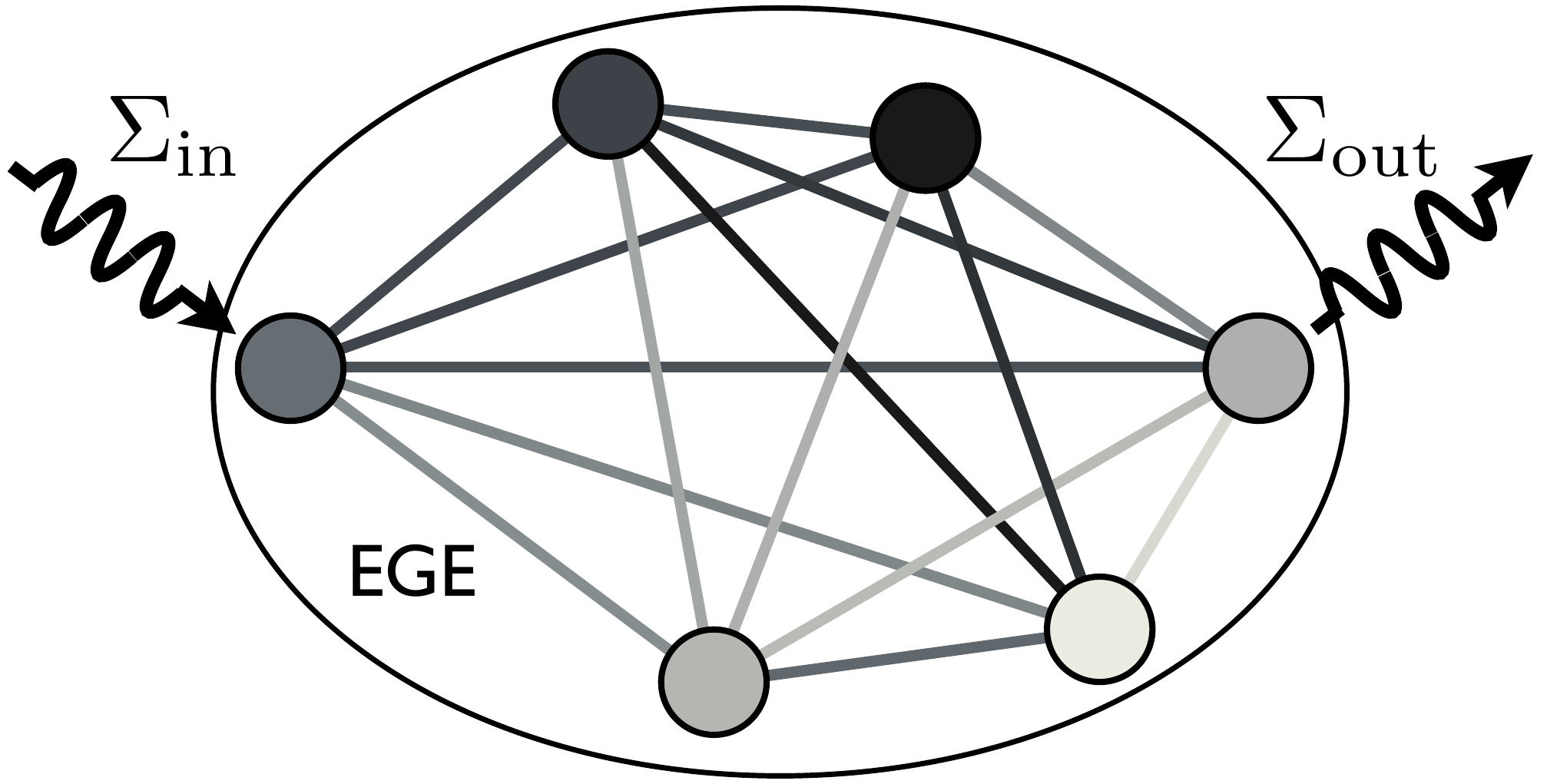}
  \includegraphics[scale=0.3]{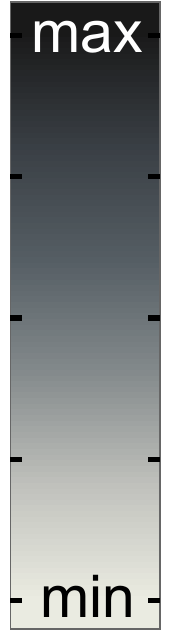}
  \includegraphics[scale=0.19]{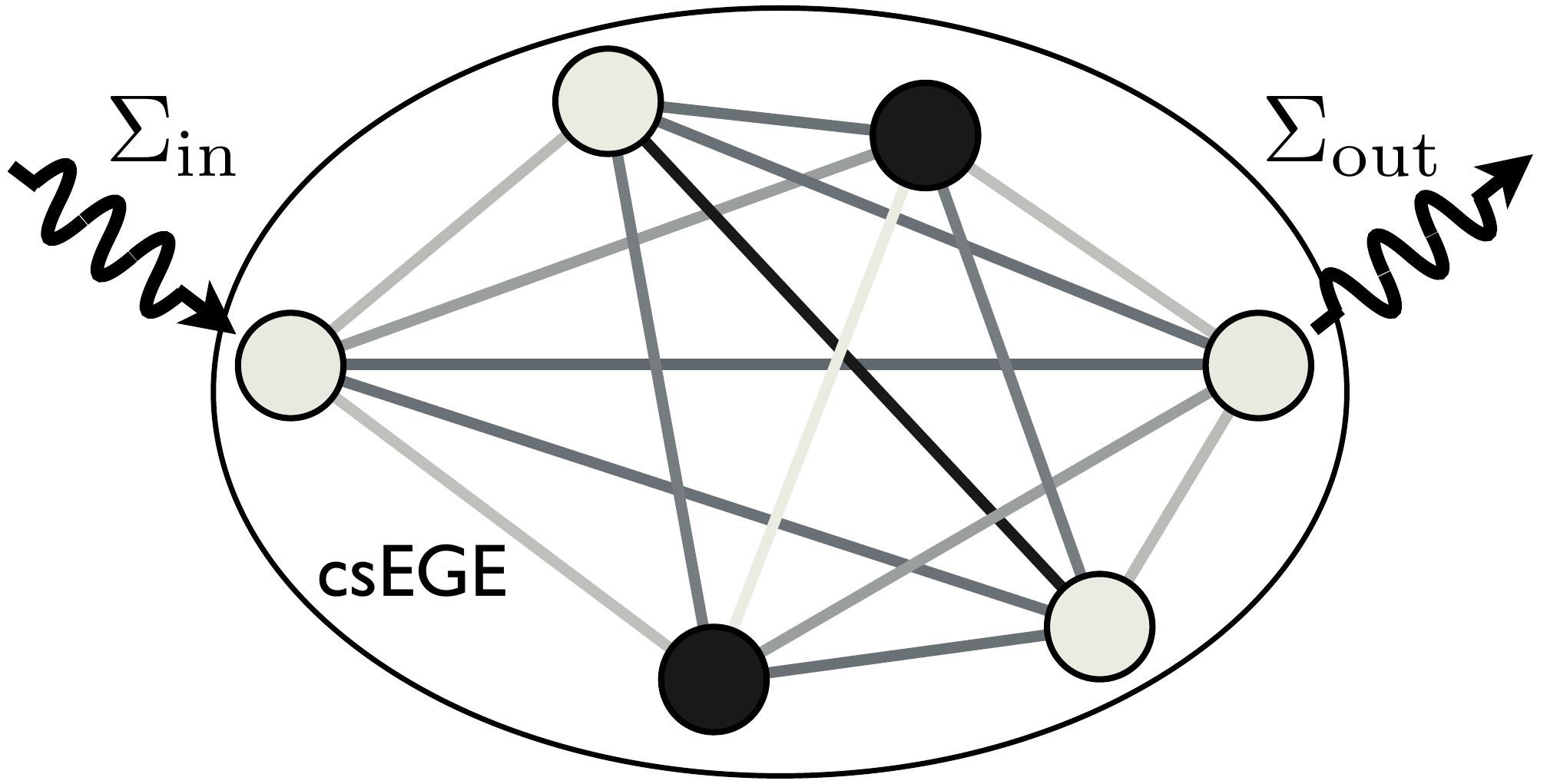}
  \caption{Graph representations of Hamiltonians from the EGE (left) and the csEGE
    (right). The shading of the nodes and links corresponds to the matrix element
    $\braket{\nu | V_k | \mu}$. The transmission between the ingoing and outgoing
    states, indicated by $\Sigma_\text{in/out}$, is shown in \Fig{1}.}
  \label{fig:2}
\end{figure}

The representation of the Hamiltonian $H_k$ can be interpreted as a graph, see \fig{2}. Each node
represents a $n$-body many-particle state $\ket{\mu} = \psi^\dagger_{n;\mu} \ket{0}$. The number of
nodes in the graph is determined by the dimension of the Hilbert space $N$. A pair of nodes is
linked, if the matrix element $\braket{\nu | V_k | \mu} \neq 0$.  The graphs generated by the
fermionic EGE model are regular graphs \cite{2001AnnPhys-BRW}, because each node has the same number
of links.

\subsection{Centrosymmetry in the embedded Gaussian ensemble}
\label{sec:Centrosymmetry}

Centrosymmetry is an important concept to enhance the efficiency \cite{2014NJP-ZMWB,
  2015ADP-OrtegaMananBenet} and, as we shall see below, transport. A symmetric $N\times N$ matrix
$A$ is centrosymmetric if $[A,J]=0$, where $J_{i,j}\equiv \delta_{i,N-j+1}$ is the {\it exchange
  matrix} ~\cite{cantoni} or, equivalently, an anti-diagonal unit matrix. One can therefore
construct a centrosymmetric matrix by imposing that a real symmetric matrix $A$ commutes with $J$.

Imposing centrosymmetry to the $k$-body embedded ensemble is
subtle. Following~\cite{2015ADP-OrtegaMananBenet}, it can be introduced either at the one-particle
level, which is the core for the definition of the $k$- and $n$-particle Hilbert spaces, at the
$k$-body level, where the actual (random) parameters of the embedded ensemble are set, or at the
$n$-body level, which defines the dynamics. The latter cases can be implemented following the
procedure described in Ref.~\cite{cantoni} (c.f. Lemmas 2(i) and 2(ii)), though it is not clear
whether we should choose the $k$ body space or the $n$ body space. Since the one-particle states are
the building blocks to construct both the $k$-body particle states and the $n$-body particle states,
we shall define centrosymmetry at the one-particle level. Note that this approach yields a
consistent treatment of more realistic situations, e.g. a system that includes a one-body
(mean-field) term and a two-body (residual) interaction, $H = H_{k=1}+H_{k=2}$.

Considering that centrosymmetry is introduced at the one-particle level, we define it by
$J_1 \ket{j} = \ket{l-j+1}$ for $j=1,2,\dots, l$, whose matrix representation in the one-body basis
is precisely the exchange matrix. For two fermions, we define
$J_2 \psi^\dagger_{2;j_1,j_2} = J_1 a^\dagger_{j_1} J_1 a^\dagger_{j_2} = -
\psi^\dagger_{2;l-j_2+1,l-j_1+1}$.
In the last equation we followed the convention that the indices are arranged in increasing order;
then, the fermionic anticommutation relations impose a global minus sign, which can be safely
ignored. This is generalized for $k$ particles as
\begin{equation}
\label{CS}
J_k \psi^\dagger_{k;j_1,\dots,j_k} = \prod_{s=1}^k J_1 a^\dagger_{j_s} =
\psi^\dagger_{k;l-j_k+1,\dots,l-j_1+1}\ ,
\end{equation}
where we have dropped any global minus sign. We note that in general the matrix $J_k$, as defined by
Eq. (\ref{CS}), may not be an exchange matrix. This follows from the possible existence of more than
one state that is mapped by $J_k$ onto itself; in this case, we shall say that $J_k$ is a {\it
  partial} exchange matrix. As an example, considering $l=4$ single-particle states and two-body
($k=2$)
interactions, the $k$-particle space has dimension $6$. In this case,
$J_2 \psi^\dagger_{2;2,3} = \psi^\dagger_{2;2,3}$ and
$J_2 \psi^\dagger_{2;1,4} = \psi^\dagger_{2;1,4}$, ignoring the minus signs mentioned above, since
under $J_1$ we have $\ket{1}\to\ket{4}$ and $\ket{2}\to\ket{3}$. Then, the entries in the $J_2$
matrix elements for these basis states are 1 in the diagonal and $J_2$ is a partial exchange
matrix. In contrast, for the case $l=4$ and $k=1,3$ the resulting matrices $J_1$ and $J_3$ are
exchange matrices.

\subsection{The nonequilibrium Green's function method for transport}
\label{sec:NEGF}

Transport is studied by means of the nonequilibrium Green's function (NEGF) method. In the
following, we discuss the physical meaning of the necessary equations. A detailed introduction and
derivations can be found in \cite{Datta1997, Datta2005, Economou2006, DiVentra2008, Cuevas2010}.

The Green's function of the system is defined as
\begin{equation}
  \label{eq:1}
  G(E)=\lr{E-H'_k}^{-1},
\end{equation}
where the \textit{effective Hamiltonian}
\begin{equation}
  \label{eq:2}
  H'_k= H_k +\Sigma_{\text{in}} +\Sigma_{\text{out}}
\end{equation}
is composed of the Hamiltonian $H_k$ of the closed quantum system (\ref{eqHam}) and the
self-energies $\Sigma_{\text{in}}$ and $\Sigma_{\text{out}}$ (defined below). A Fourier transform
from the energy to the
time domain shows that the matrix elements of the Green's function $G_{i,j}(t)$ describe the
response of the state $j$ at time $t$ after a $\delta(t)$-excitation of the state $i$ at time
$t=0$~\cite{Datta1997, Datta2005}. Hence, the Green's function describes the propagation of excitation
through the many-body states of the quantum system.

The self-energy matrix elements,
\begin{eqnarray}
  \label{eq:3}
        {\Sigma_\text{in}}_{r,s} & = & - \I \eta \, \delta_{r,\text{in}} \delta_{r,s}, \\
        {\Sigma_\text{out}}_{r,s} & = & - \I \eta \, \delta_{r,\text{out}} \delta_{r,s}, \nonumber
\end{eqnarray}
indicate the ingoing and outgoing states, where the excitation enters and leaves the system from the
environment, see \fig{2}. The transport is studied between these two states. For the self-energies
we have used the so called wide-band approximation \cite{Verzijl2013}, where the density of states
of the environment is constant ($\sim \eta$) for all considered energies. This is certainly not a
microscopic exact model of the environment, but it captures the essential effect: The self-energies
are a non-Hermitian modification of the Hamiltonian, so in general its eigenvalues are complex. The
imaginary part of the eigenvalues indicates the finite lifetime or energy-broadening of the states
due to the coupling to the environment. The broadening is parametrized by $\eta$ for which we will
take in the following $\eta=1$.

The self-energies $\Sigma_{\text{in}}$ and $\Sigma_{\text{out}}$ can break the centrosymmetry of the
system. We shall
consider only self-energies under which the system Hamiltonian $H_k$ and the effective Hamiltonian
$H'_k$ are centrosymmetric. In the following, we chose as the ingoing state
$\ket{1} = \ket{1,1,\dots,1,0,\dots,0}$, where all the fermions are shifted to the left. As the
outgoing state we take $\ket{N} = \ket{0,0,\dots,0,1,\dots,1}$, where all the fermions are shifted
to the right. These states are clearly related to each other by centrosymmetry, which in turn
implies the centrosymmetry of the effective Hamiltonian if $H_k$ is centrosymmetric.

The transmission probability between the ingoing ingoing and outgoing states is given by
\begin{equation}
  \label{eq:4}
  T(E)= 4 \text{Tr} \bigl(\Im{\Sigma_{\text{in}}} G \Im{\Sigma_{\text{out}}} G^\dagger \bigr)
  = |2G_{\text{in}, \text{out}}|^2.
\end{equation}
Note that in the last equality, we have used the fact that the self-energy matrices \eq{3} have only
one non-vanishing matrix element. This equation, which is known also as the Caroli formula
\cite{Caroli1971}, has a clear physical interpretation. As the Green's function describes the
propagation of excitations in the system, the transmission is just the absolute square of the matrix
element of the Green's function, which connects the ingoing and outgoing states.

The total current through the system is given by integrating the transmission over all energies,
i.e.
\begin{equation}
  \label{eq:5}
  I = \int_{-\infty}^\infty \d E \, T(E).
\end{equation}

\section{Results and discussion}
\label{sec:ResultsDiscussion}

\subsection{Ensemble averaged transmission and current distributions}

\begin{figure*}[htb]
  \includegraphics[scale=0.28]{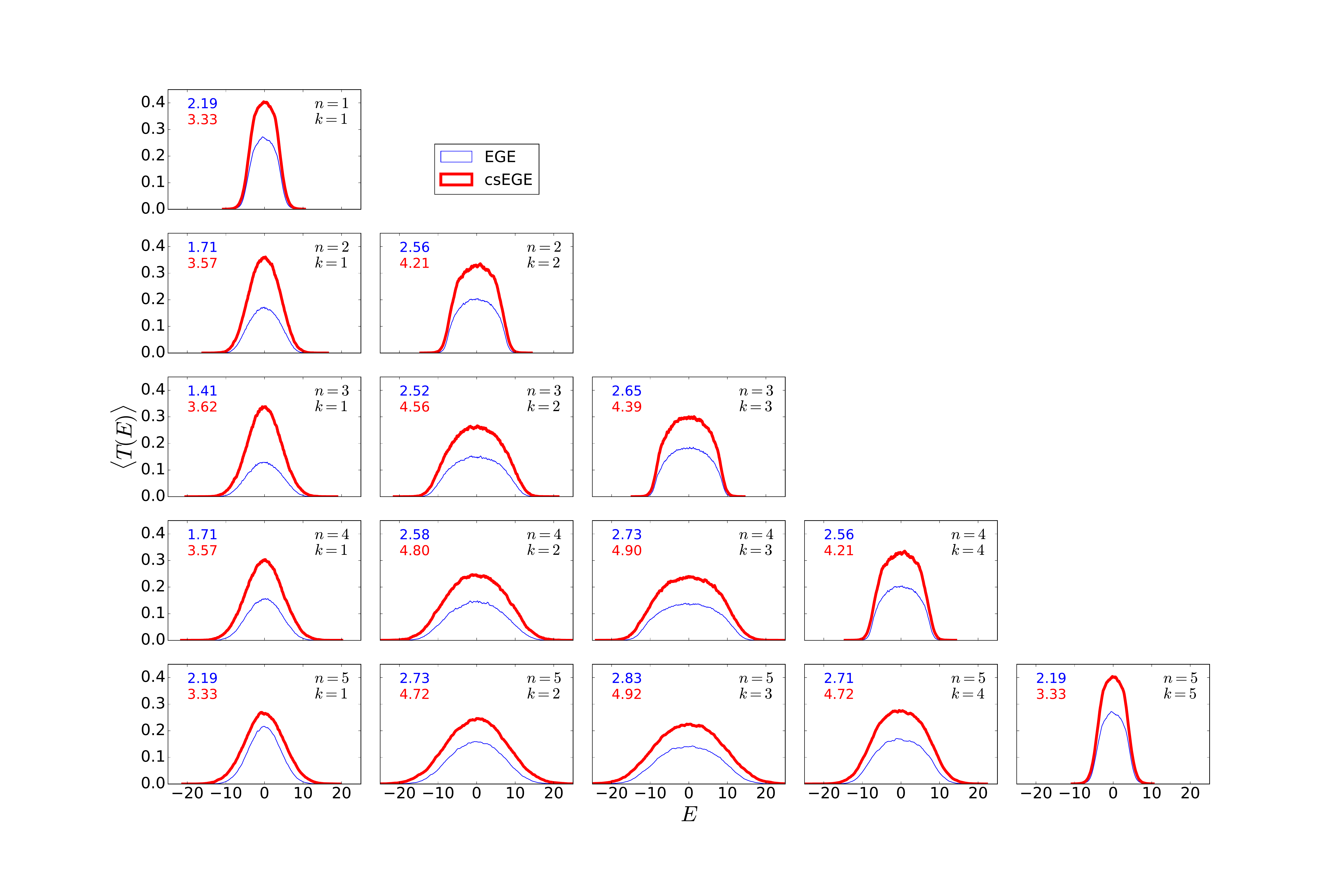}
  \vspace{-2mm}
  \caption{(color online) Ensemble averaged transmission $\braket{T(E)}$ as a function of the energy
    $E$ for a system of $l=6$ single-particle states. Each column has fixed value of $k$, for
    $k=1,2,\dots,n$, while each row corresponds to a fixed value of $n$, for $n=1,2,\dots,l-1$. The
    ensemble consists of $10^4$ realizations. The results corresponding to the EGE are displayed by
    the blue-thin curves, while the red-thick curves illustrate the csEGE results. Imposing centrosymmetry
    increases considerably the ensemble averaged transmission for all energies.}
  \label{fig:3}
\end{figure*}

In the following, we illustrate our results for a system consisting of $l=6$ single-particle states
which are occupied with $n=1,2\dots l-1$ fermions and coupled by $k$-body ($k = 1,2,\dots,n$)
interactions. One could argue that interactions between $k \sim n$ particles are averaged out and
not relevant. However, in the case $k = n$ the Hamiltonian (\ref{eqHam}) is identical to the
Gaussian orthogonal ensemble (GOE) \cite{guhr}, which has minimum
information~\cite{1968Balian}. Moreover, transport in biomolecules takes place on a sub-picosecond
time-scale \cite{Fleming2011}, where correlations between many particles can be relevant. This
justifies to address all rank of interactions. Our results remain valid for other small system
sizes, as shown in the Appendix for systems consisting of $l=8$ and $l=10$ single-particle
states. Unless stated otherwise, for each concrete set of parameters, we have calculated an ensemble
of $10^4$ realizations with and without centrosymmetry being imposed.

We have already seen in \fig{1} that centrosymmetry enhances significantly the coherent transmission
$T(E)$ through the system. The ensemble averaged transmission $\braket{T(E)}$ is displayed in
\fig{3}, where all combinations of $n$ and $k$ are shown. We observe that in the case of
centrosymmetric embedded Gaussian random matrix ensemble (csEGE) the transmission is for all
energies higher than for the non-centrosymmetric Gaussian ensemble (EGE), i.e.
$ \braket{T_\text{csEGE}(E)} > \braket{T_\text{EGE}(E)} $. In both cases the spectral span of the
transmission, i.e., the width of the conduction band, is maximal for $k \sim n/2$ and increases with
$n$. That is, the system is conductive for a wider range of energies. The ensemble averaged
transmission is peaked around the center of the conduction band at $E=0$.

In particular, for fixed $n$ (along a row in \fig{3}) maximal values of the transmission are
attained at $k=1$ and $k=n$ for the EGE as well as for the csEGE. The shape of the maxima differs
in the two cases; for $k=1$ the maximum is strongly peaked, whereas for $k=n$ it is broader and
rounded. This behavior can be observed also for larger systems, see the Appendix, though the values
for $k=1$ are slightly larger than for $k=n$. These cases are of interest because transport is
typically more efficient in a narrow energy band around $E=0$. This effect is most pronounced for
the extrema at $k=n=1$ and $k=n=5$.

\begin{figure*}[t]
  \includegraphics[scale=0.28]{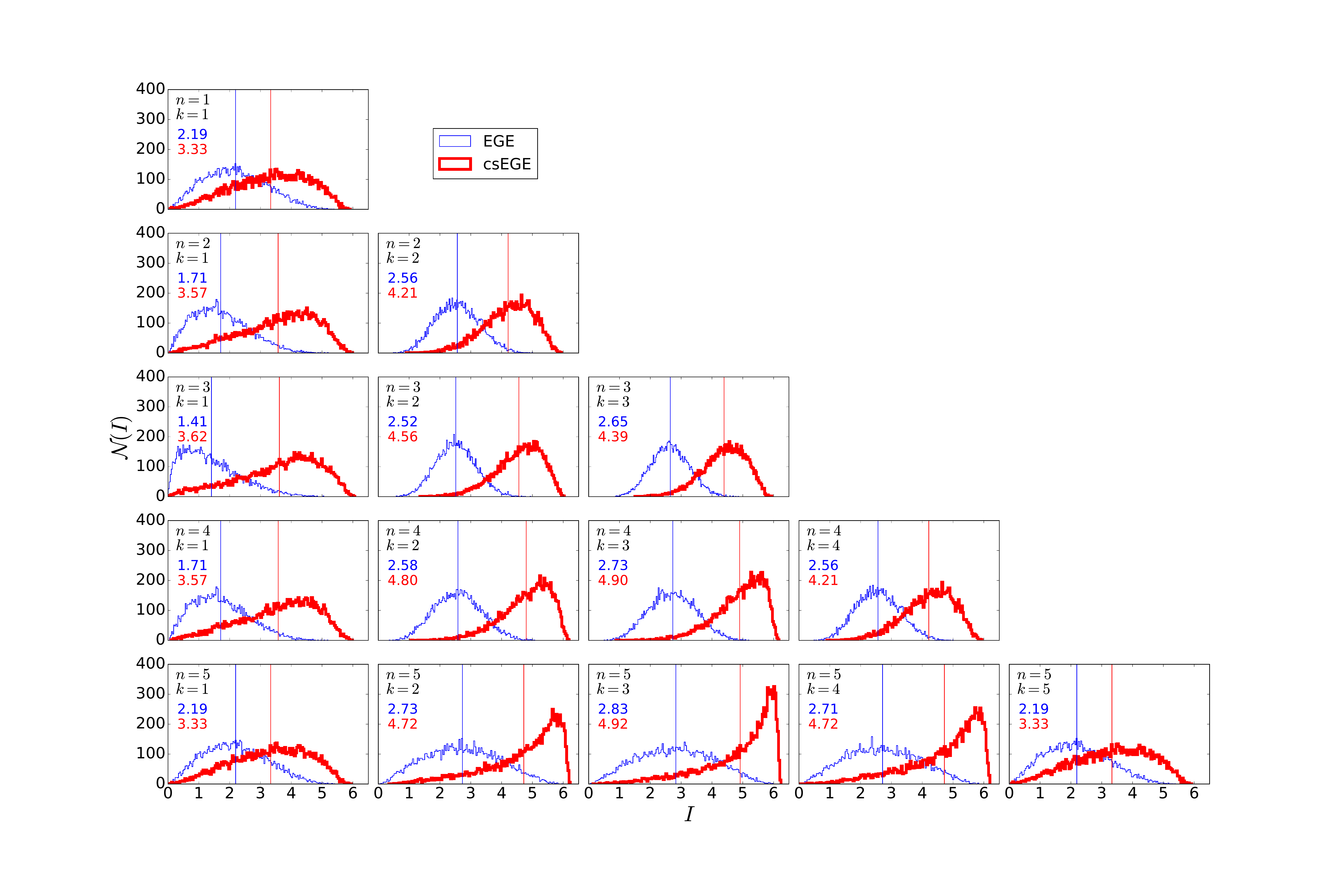}
  \caption{(color online) Frequency histogram of the current for $10^4$ realizations for the EGE
    (blue-thin) and the csEGE (red-thick). The arrangement of the figures is the same as in \fig{4}.  The
    average current $\braket{I}$ is indicated by the vertical lines and their values are
    shown in the insets. The current is maximal if the system is almost filled
    $n = l-1$ and the rank of interaction between the particles $k\sim n/2$.}
  \label{fig:4}
\end{figure*}

In \fig{4} we present the frequency histograms $\mc{N(I)}$ of the current, calculated by means of
\eq{5}. We observe that the average current $\braket{I}$, whose values are included in the insets
and are illustrated by the vertical lines, is enhanced significantly when centrosymmetry is
imposed. This trend is independently from the actual value of the parameters $(n,k)$. Moreover, the
average current is maximal if the system is almost filled, i.e.  $n = l-1$, and the rank of
interaction is $k \sim n/2$.  These statements also apply to the mode (i.e. the position of the
maximum) of the current distributions.  Note that for larger systems (see the Appendix) the average
current is maximal for $n$ close, but not identical, to $l-1$.  These results for stationary
transport are fully consistent with our previous results \cite{2015ADP-OrtegaMananBenet}, where the
dynamic propagation of states was addressed. The effects of centrosymmetry are thus present in
time-dependent quantities \cite{2013PRL-walschaers, 2015ADP-OrtegaMananBenet, 2015PRE-walschaers}
and also in stationary transport properties.

\begin{figure*}[t]
  \includegraphics[scale=0.38]{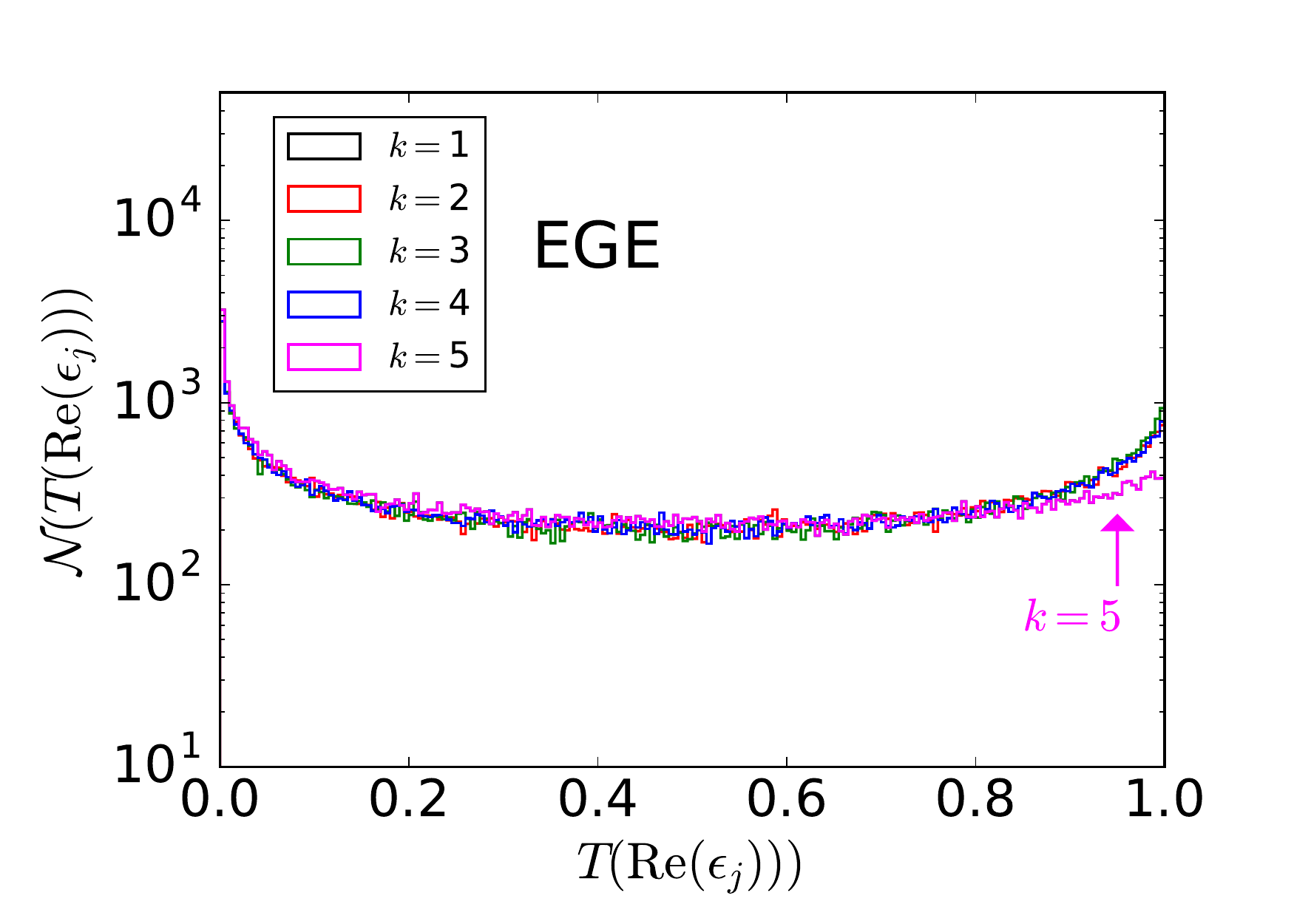}
  \includegraphics[scale=0.38]{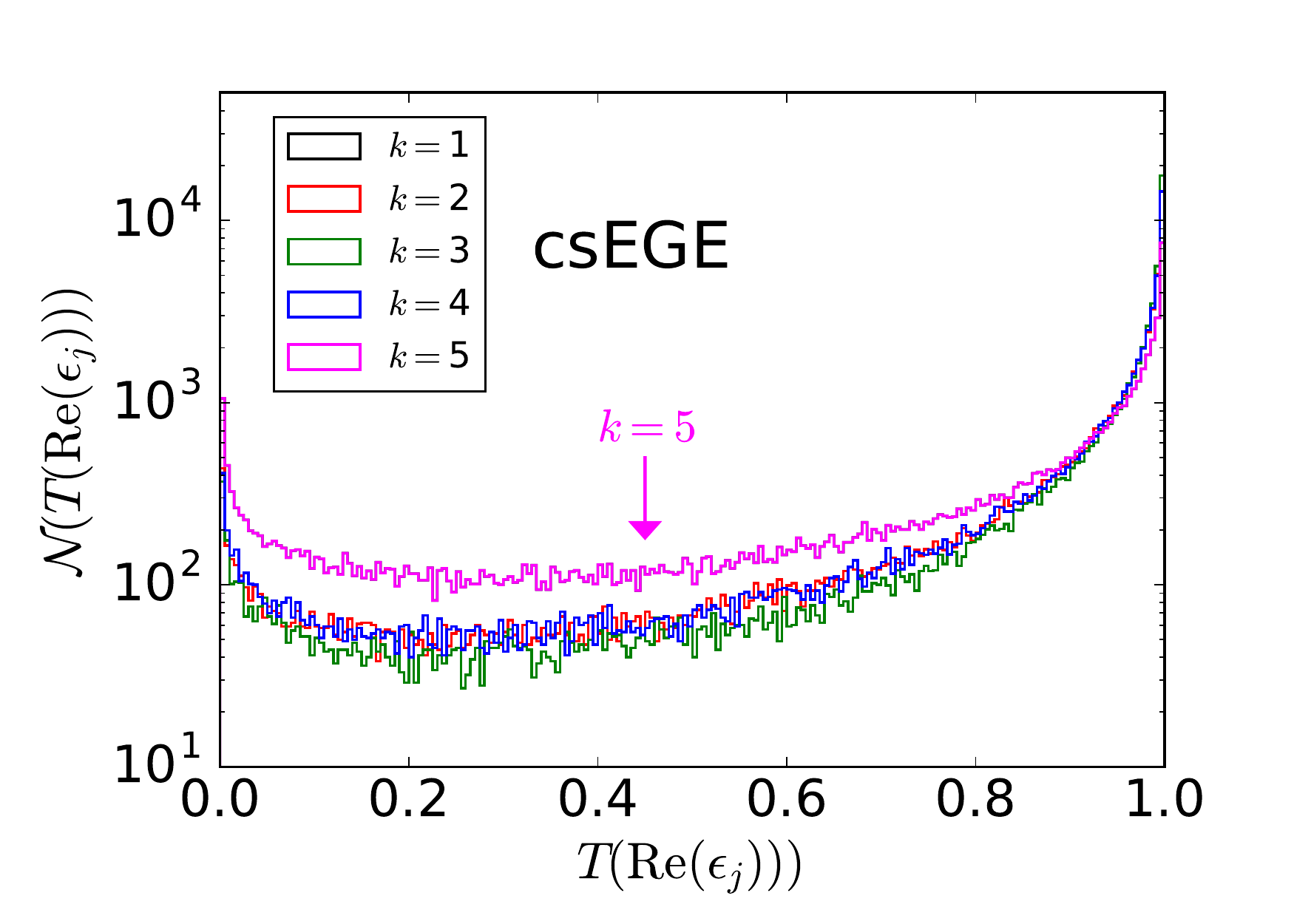}\\[-2mm]
  \includegraphics[scale=0.38]{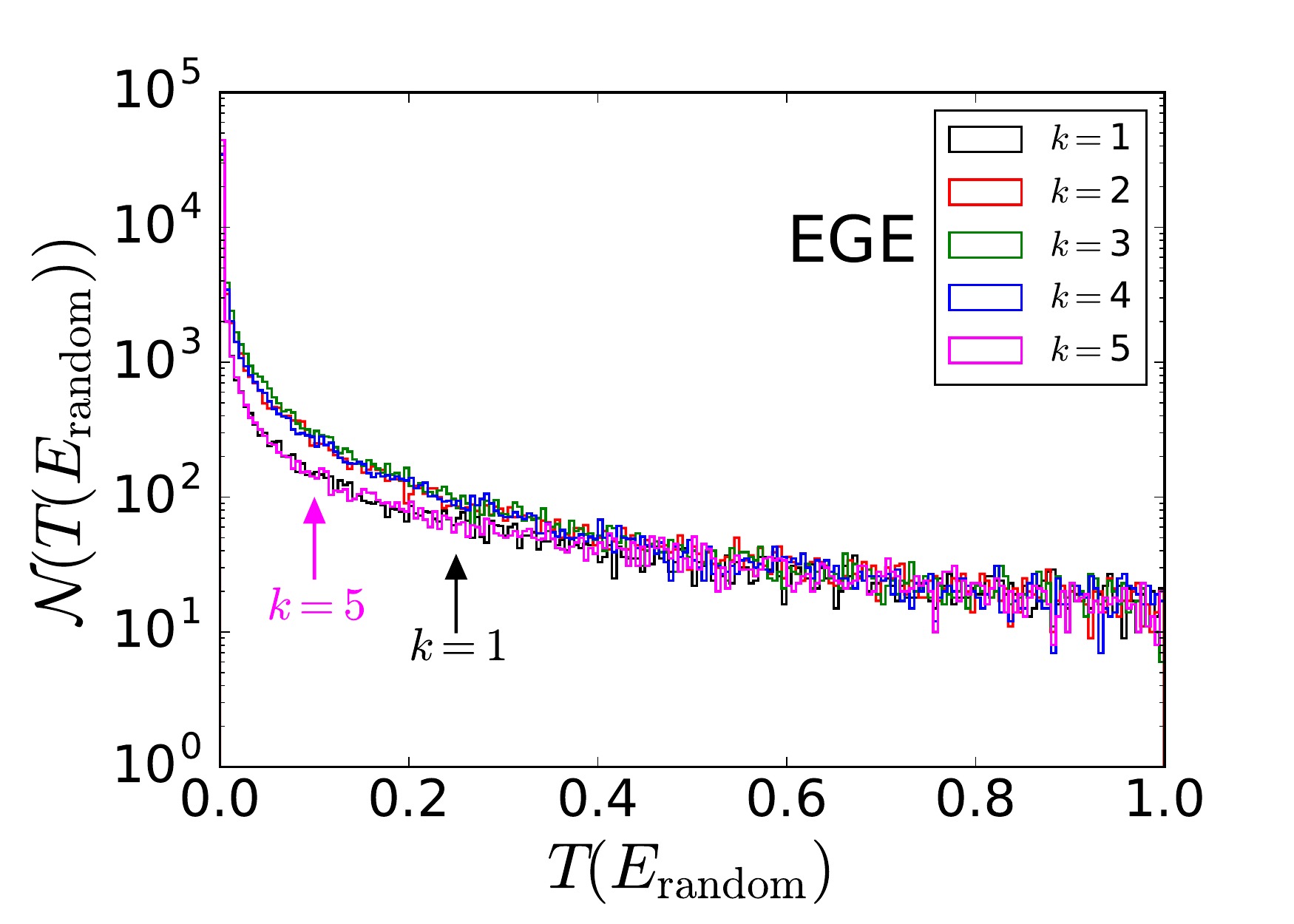}
  \includegraphics[scale=0.38]{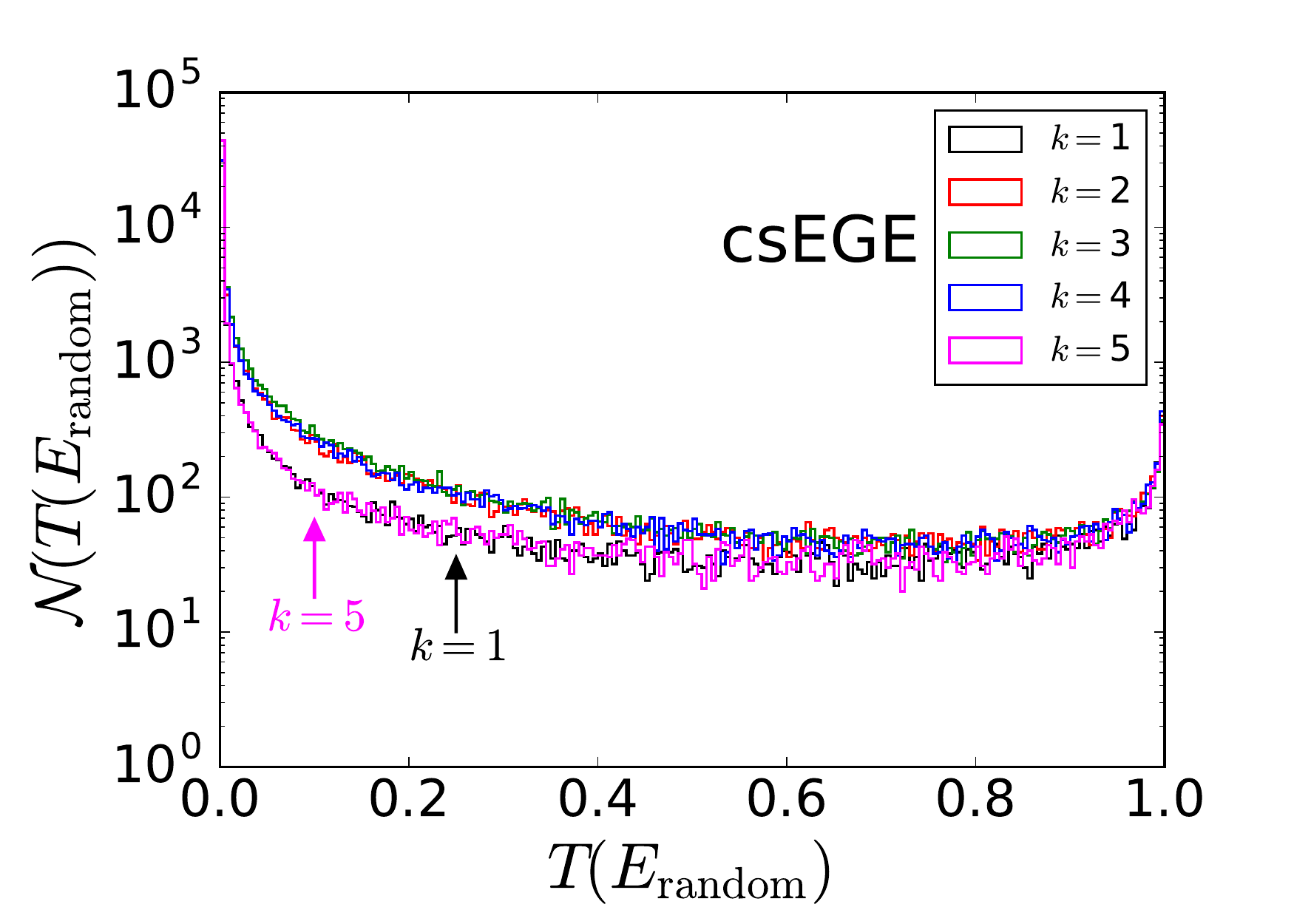}
  \caption{(color online) Top row: Histogram of the transmission evaluated at the real part of the
    complex eigenenergies of $H'_k$, where the resonances are located approximately. The histogram
    confirms our observation from \fig{1} that centrosymmetry generates many resonances of optimal
    transmission $T(\Re{\eps_j})= 1$. Bottom row: The histogram of the transmission at random
    energies confirms this property and furthermore, shows the general trend that centrosymmetry
    enhances the transmission for all energies, see \fig{3}. Note that the vertical axes are in
    logarithmic scale.}
\label{fig:5}
\end{figure*}

As we consider finite systems of $l$ states occupied by $n$ particles, one may look for
particle-hole symmetries in the system. The Hamiltonian $H_k$ may describe $n$ particles as well as
$l-n$ holes; yet, the embedded ensemble in general do not display that symmetry, see
\cite{2001AnnPhys-BRW}.  Briefly, applying the particle-hole transformation to $H_k$, the result is
a Hamiltonian that consists of the sum of ranks $0, 1, \dots, k$ in the hole representation, instead
of a single term of rank $k$.  As for $k=n$ the $H_k$ are taken from the GOE, we observe
particle-hole symmetry in these cases, i.e. results for the parameters $(n,k=n)$ and $(l-n, k=l-n)$
are identical. This can be seen clearly in the ensemble averaged transmission as well as in the
frequency histogram of the total current, see \fig{3} and \fig{4} as well as \fig{12} - \fig{15} in
the Appendix. Further symmetries can be observed only in the distribution of the total current
(Figure 4), where we observe numerically that the cases $(n,k=1)$ and $(l-n,k=1)$ are identical.

As illustrated in \fig{1}, we observe that centrosymmetry yields many resonance peaks with perfect
transmission, i.e. $T=1$; for EGE we may find some perfect transmission resonances, but it is not
the typical case. Denoting by $\eps_j$ the eigenvalues of $H'_k$, these resonances are located at
energies $E\approx \Re{\eps_j}$. This motivates us to study in \fig{5}
the statistics of the transmission at at these energies $\Re{\eps_j}$ (top row), and compare them with the transmission at
random energies (bottom row) for the EGE and the csEGE in the case $n=l-1$ and all possible values
of $k$. These histograms confirm our observation about \fig{1} that centrosymmetry generates many
resonances with perfect transmission. Note that perfect transmission is also observed when the
transmission is evaluated at random values of the energy (within the conductance band); yet, the
relative frequency is higher by about two orders of magnitude for the csEGE. We also note that
there is a weak dependence on $k$, such that $k\sim n/2$ dominates for larger values of the
transmission. These results show the general trend that centrosymmetry enhances the transmission for
all energies, see \fig{3}, which implies a higher total current.

\begin{figure}
  \includegraphics[scale=0.35]{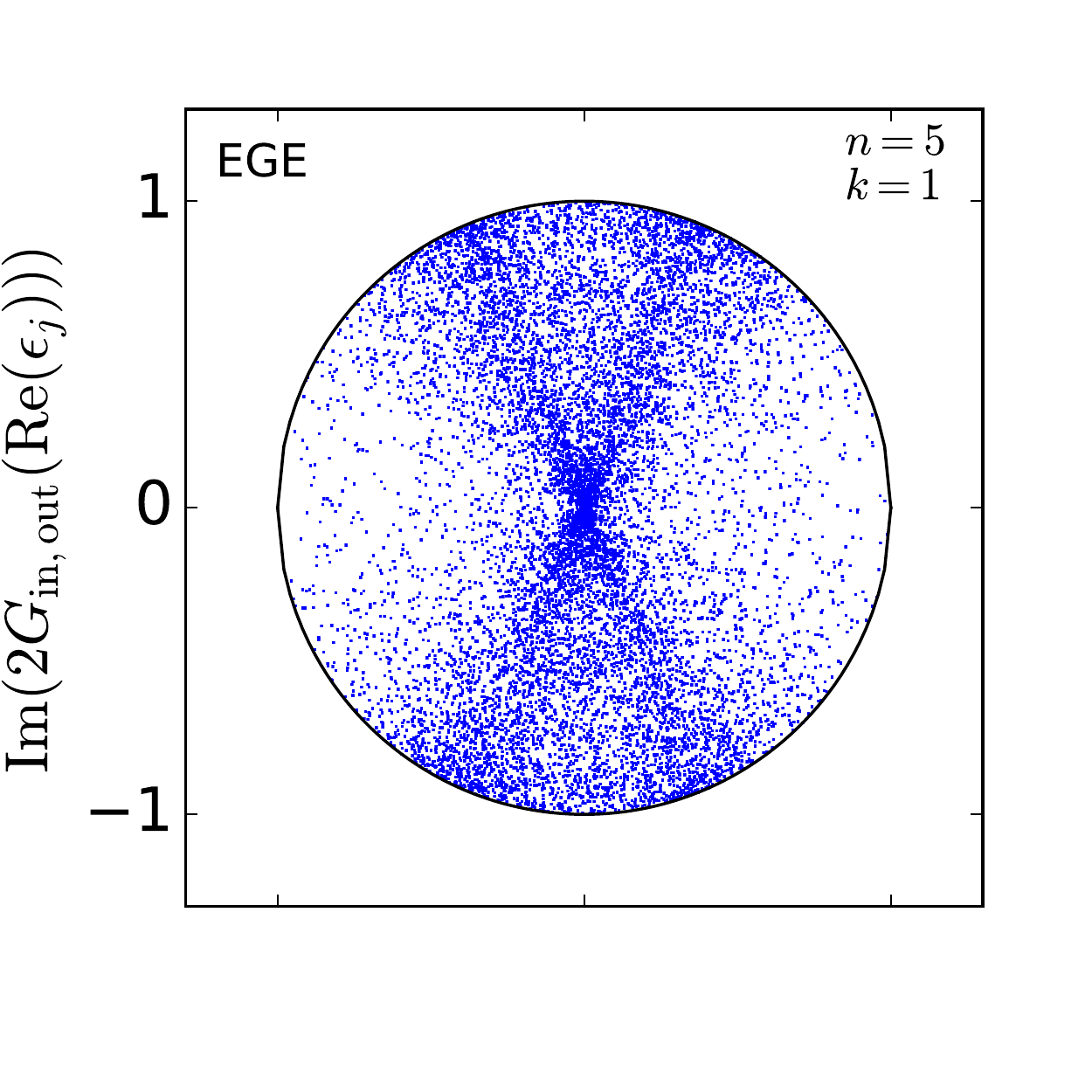}
  \hspace*{-5mm}
  \includegraphics[scale=0.35]{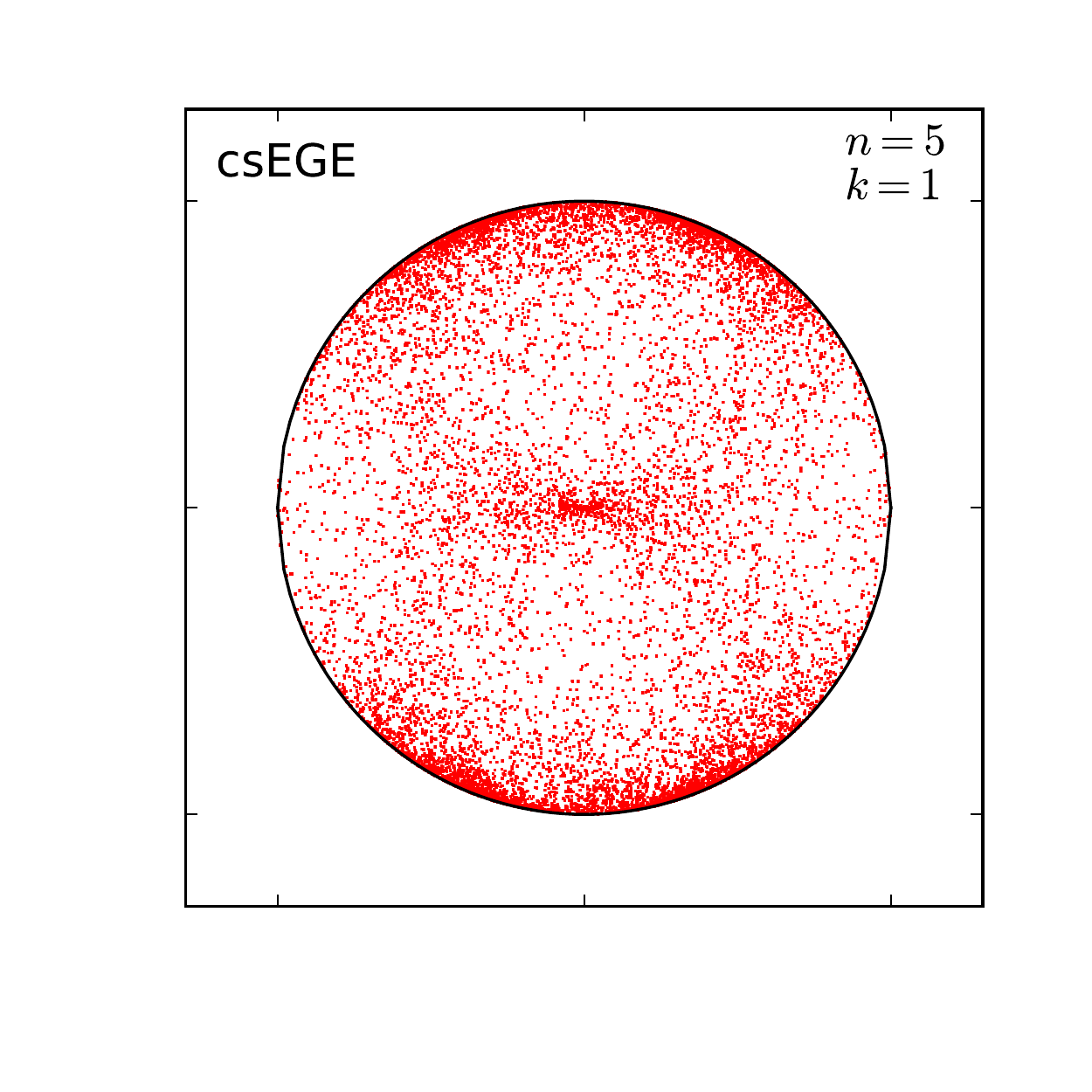}\\[-4mm]
  \includegraphics[scale=0.35]{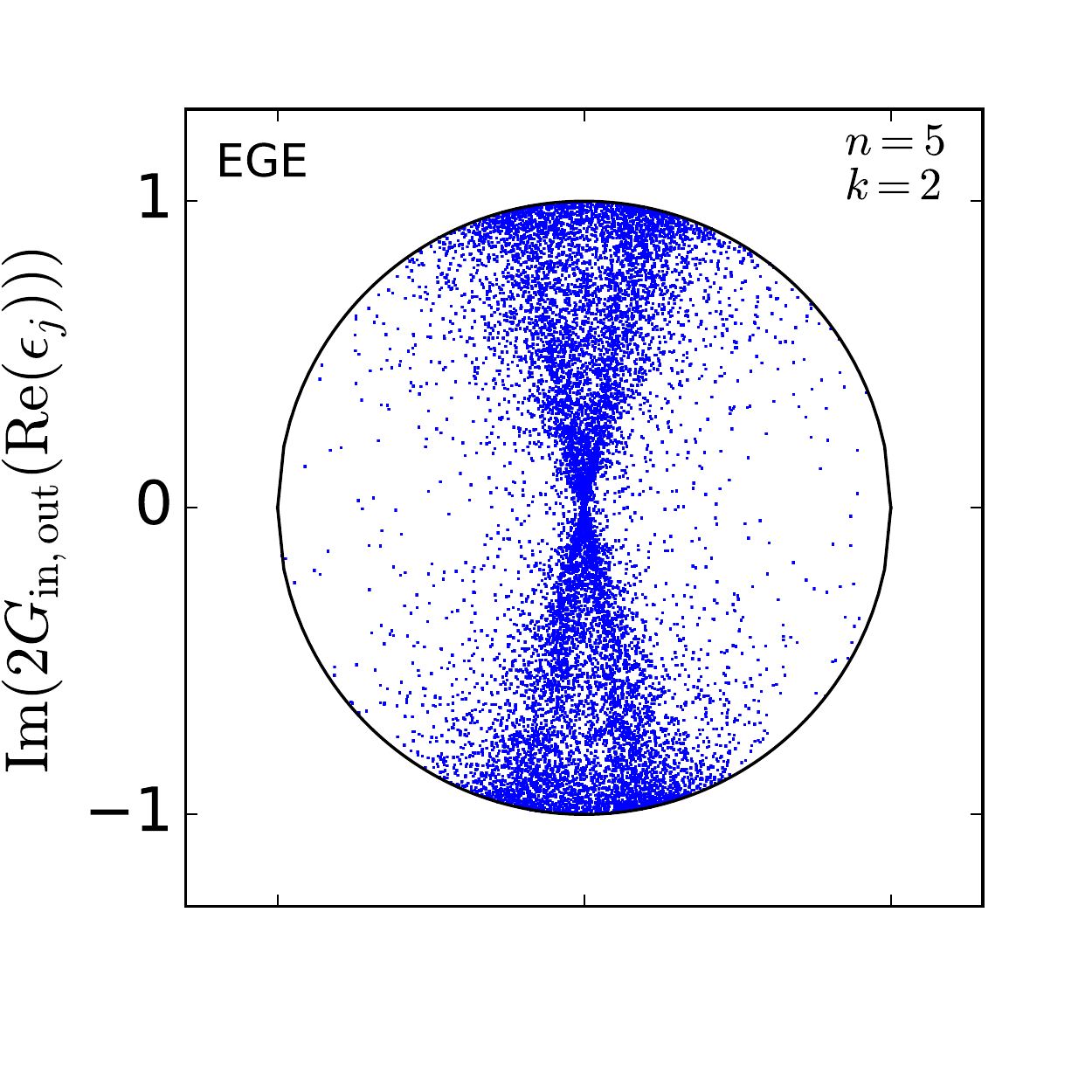}
  \hspace*{-5mm}
  \includegraphics[scale=0.35]{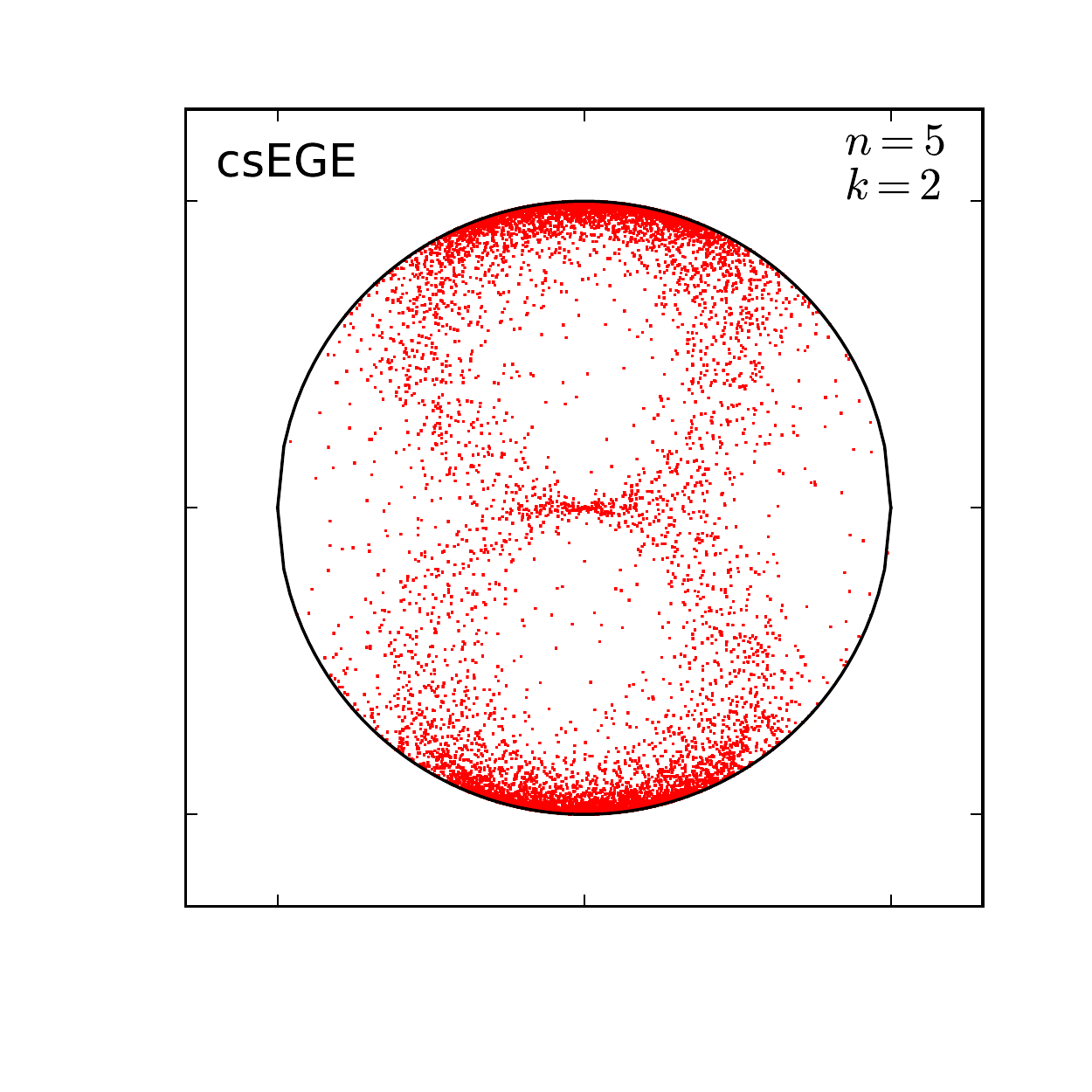}\\[-4mm]
  \includegraphics[scale=0.35]{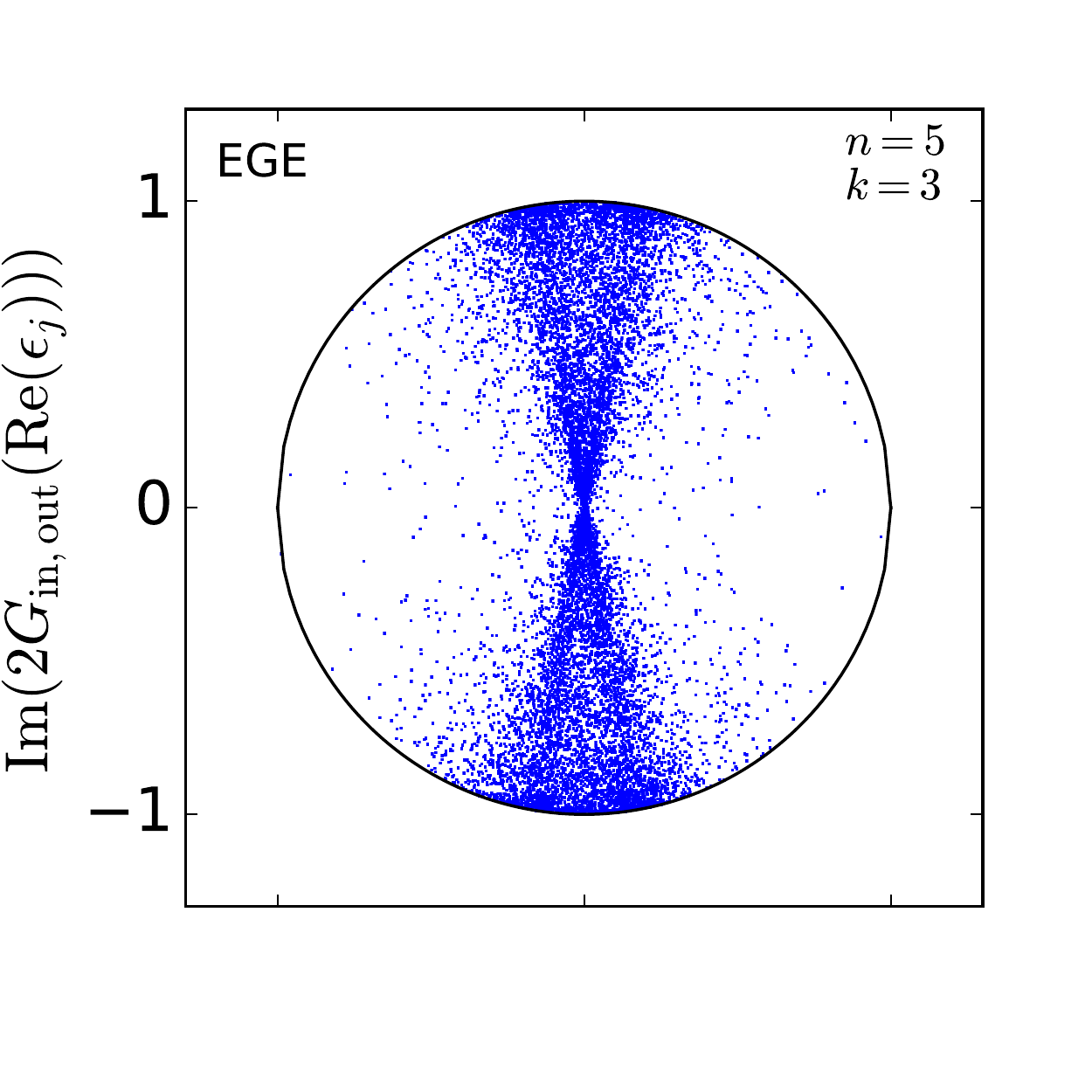}
  \hspace*{-5mm}
  \includegraphics[scale=0.35]{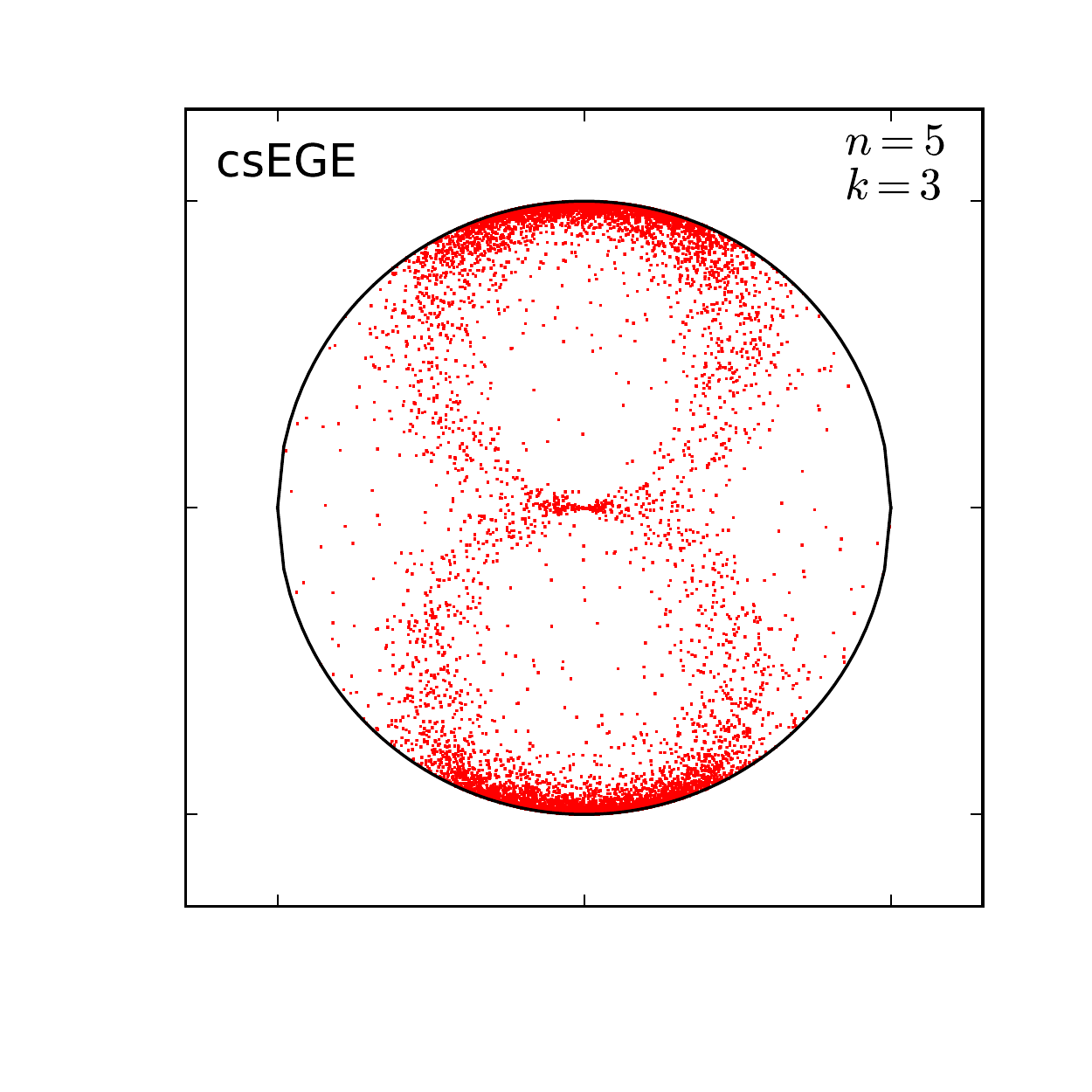}\\[-4mm]
  \includegraphics[scale=0.35]{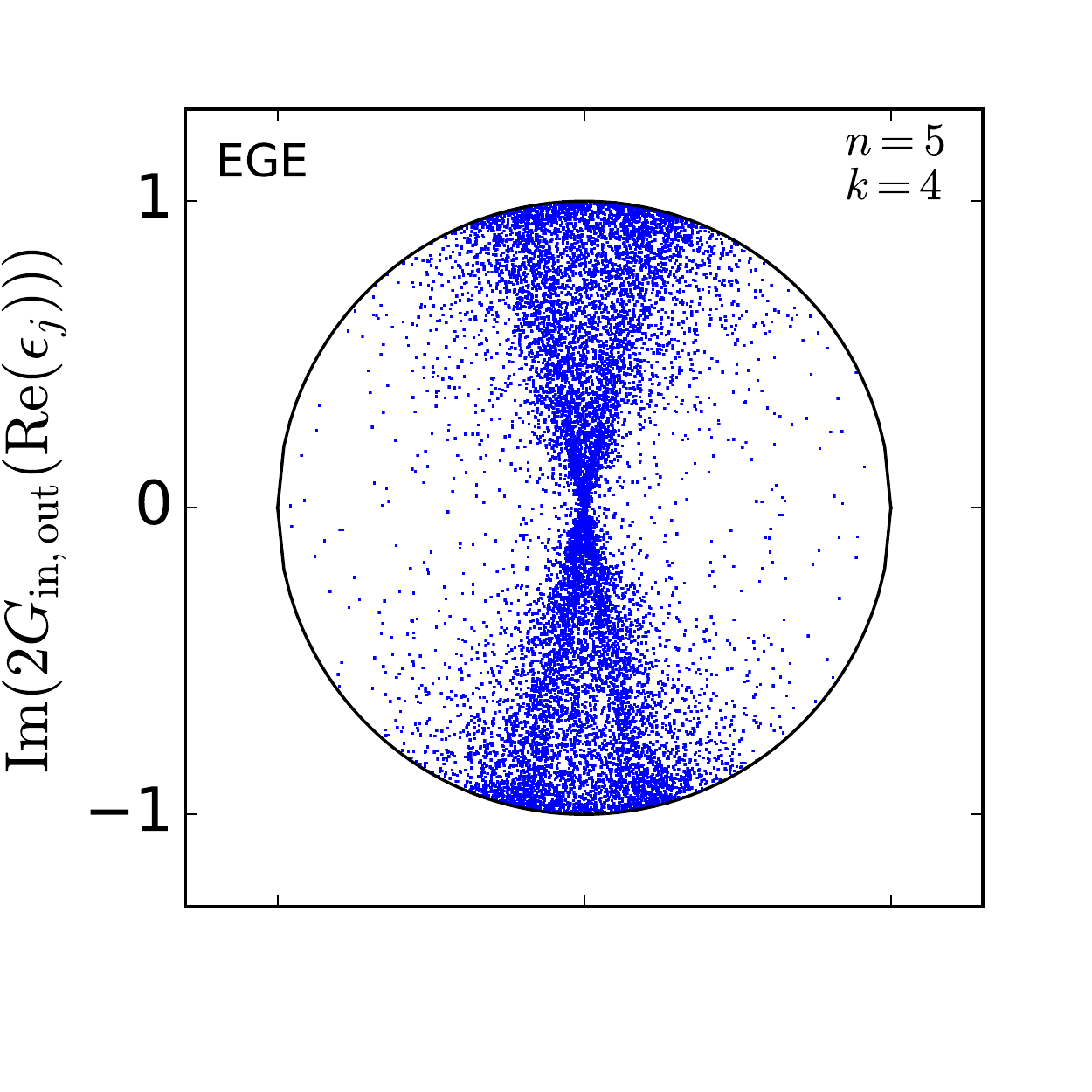}
  \hspace*{-5mm}
  \includegraphics[scale=0.35]{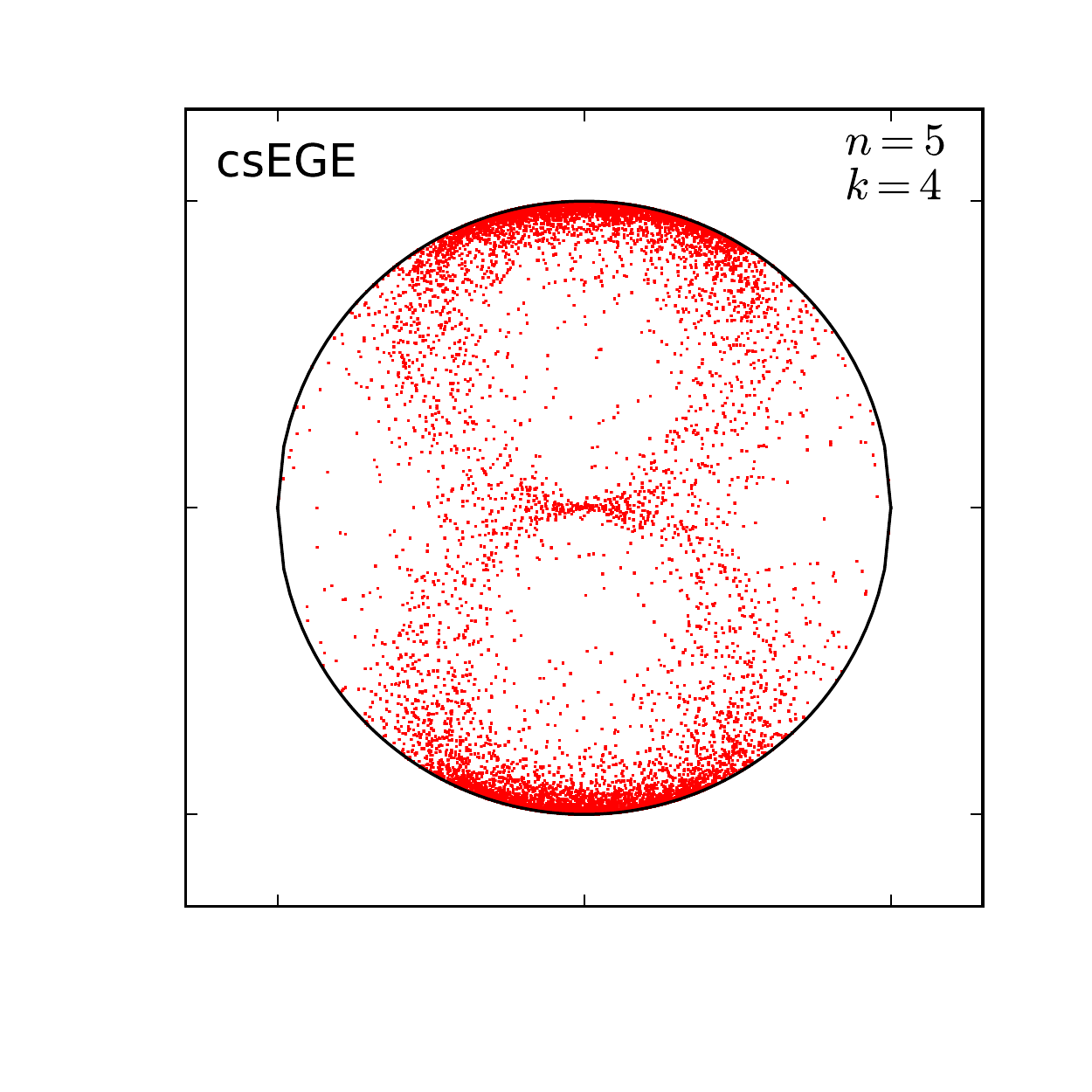}\\[-4mm]
  \includegraphics[scale=0.35]{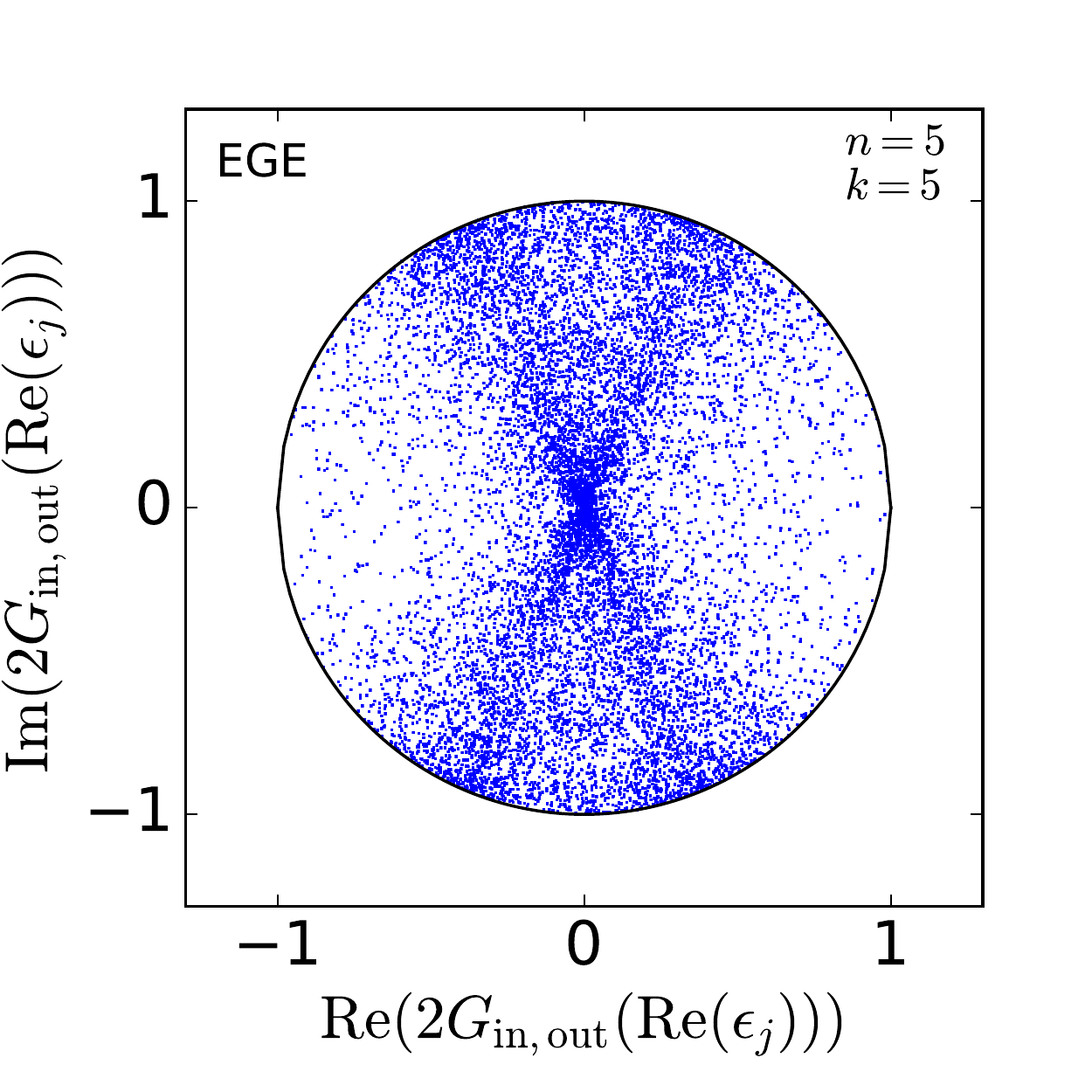}
  \hspace*{-5mm}
  \includegraphics[scale=0.35]{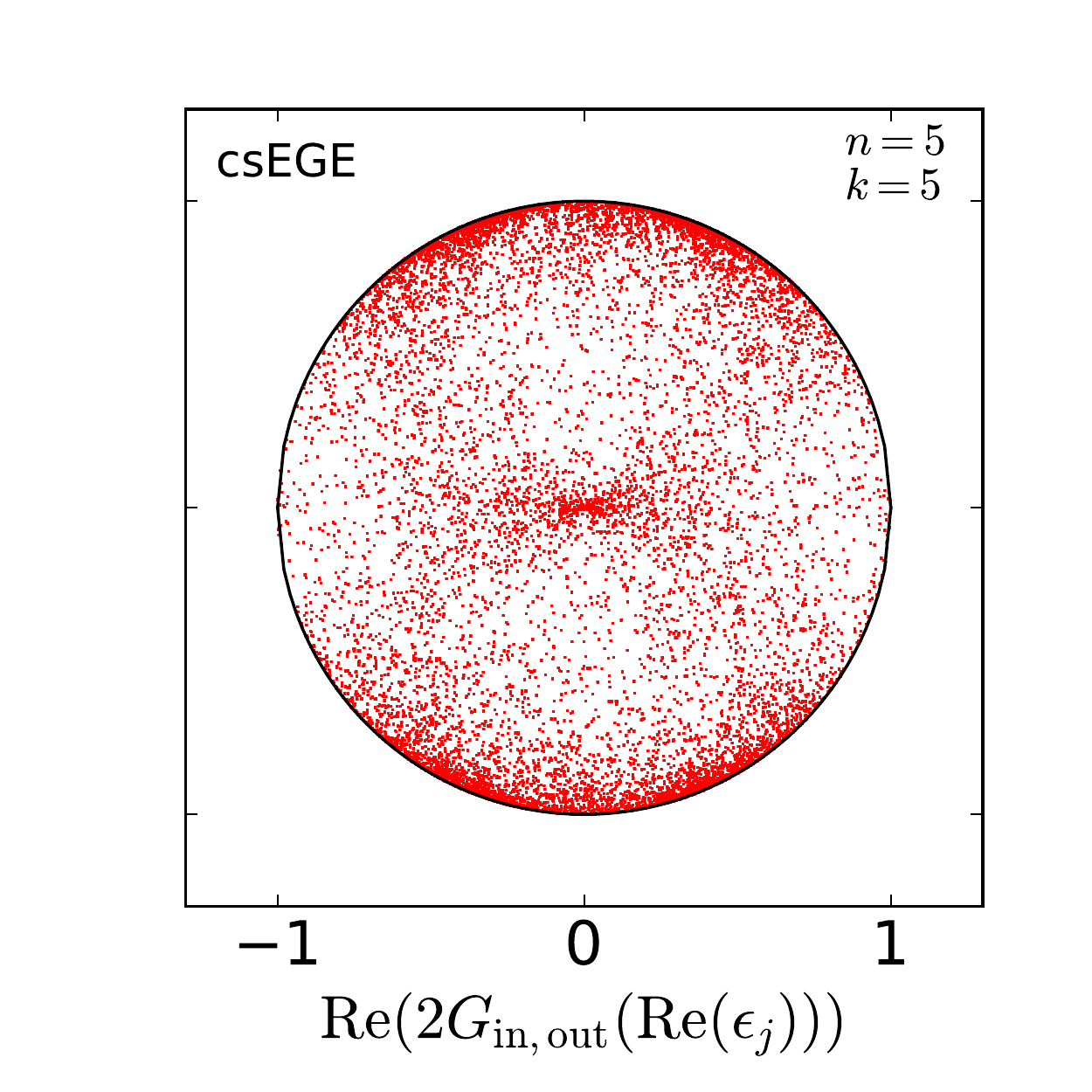}
  \caption{(color online) Distribution of $ 2G_\text{in,out}(\Re{\eps_j}) $ in the complex plane for
    2000 realizations. Strong correlations between the real and imaginary part are observed in both
    cases. In the case of EGE (left column), data points are concentrated around the origin, which
    corresponds to transmission resonances of low conductance. For the csEGE (right column), the
    data points are accumulated on the poles, which corresponds to perfect transmission.}
  \label{fig:6}
\end{figure}

In \fig{6} we show the distribution of $2G_\text{in,out}(\Re{\eps_j})$ in the complex plane for
$n=l-1$ and all values of $k$. This quantity is of interest since its modulus squared gives the
transmission through the system at $E=\Re{\eps_j}$, see \eq{4}. As the transmission is bounded to
values equal or less than 1, the data points are distributed inside the unit circle. Strong
correlations between the real and imaginary part are found for both EGE and csEGE. For the EGE
(left column), the data points are clustered around the origin, which corresponds to transmission
resonances with low conductance. In contrast, in the case of csEGE (right column), the data points
display an accumulation on the boundary of the unit circle, around the poles, corresponding to
resonances of optimal transmission. We can also see that this accumulation is larger for
$k \sim n/2$.

\subsection{Statistics of the spectral decomposition of the transmission}
\label{sec:understanding}

In order to have more insight into the effects induced by centrosymmetry on transport, we use the
spectral decomposition of the Green's function. Then, the transmission is expressed as
\begin{equation}
  \label{eq:Tdecomp}
  T(E)= \Abs{2G_{\text{in,out}}(E)}^2= \Abs{\sum_{j=1}^N \frac{\Upsilon_j}{E-\eps_j}}^2,
\end{equation}
where
\begin{equation}
\label{eq:Ups}
\Upsilon_j \equiv 2\phi_{j,\text{in}} \, \phi_{j,\text{out}}.
\end{equation}
Here, the $\phi_{j,\text{in}}$ or $\phi_{j,\text{out}}$ are the in/out components of the
$j$th eigenfunction of the
effective Hamiltonian $H'_k$. Note that $H'_k$ is non-Hermitian but it has the property
${H'_k}^\dagger={H'_k}^*$. In this case the eigenstates can be chosen in such a way that they
fulfill the orthogonality relation $\braket{\phi_i | \phi_j}= \delta_{ij}$ and the completeness
relation $1= \sum_j \ket{\phi_j} \bra{\phi_j}$, which have been used in \eq{Tdecomp}.

In the following, we will focus on the case $n = l-1 = 5$, which corresponds to the optimal case in
terms of transport; see \fig{3} and \fig{4}. Our results hold also for other $n$.

\begin{figure}
  \includegraphics[scale=0.35]{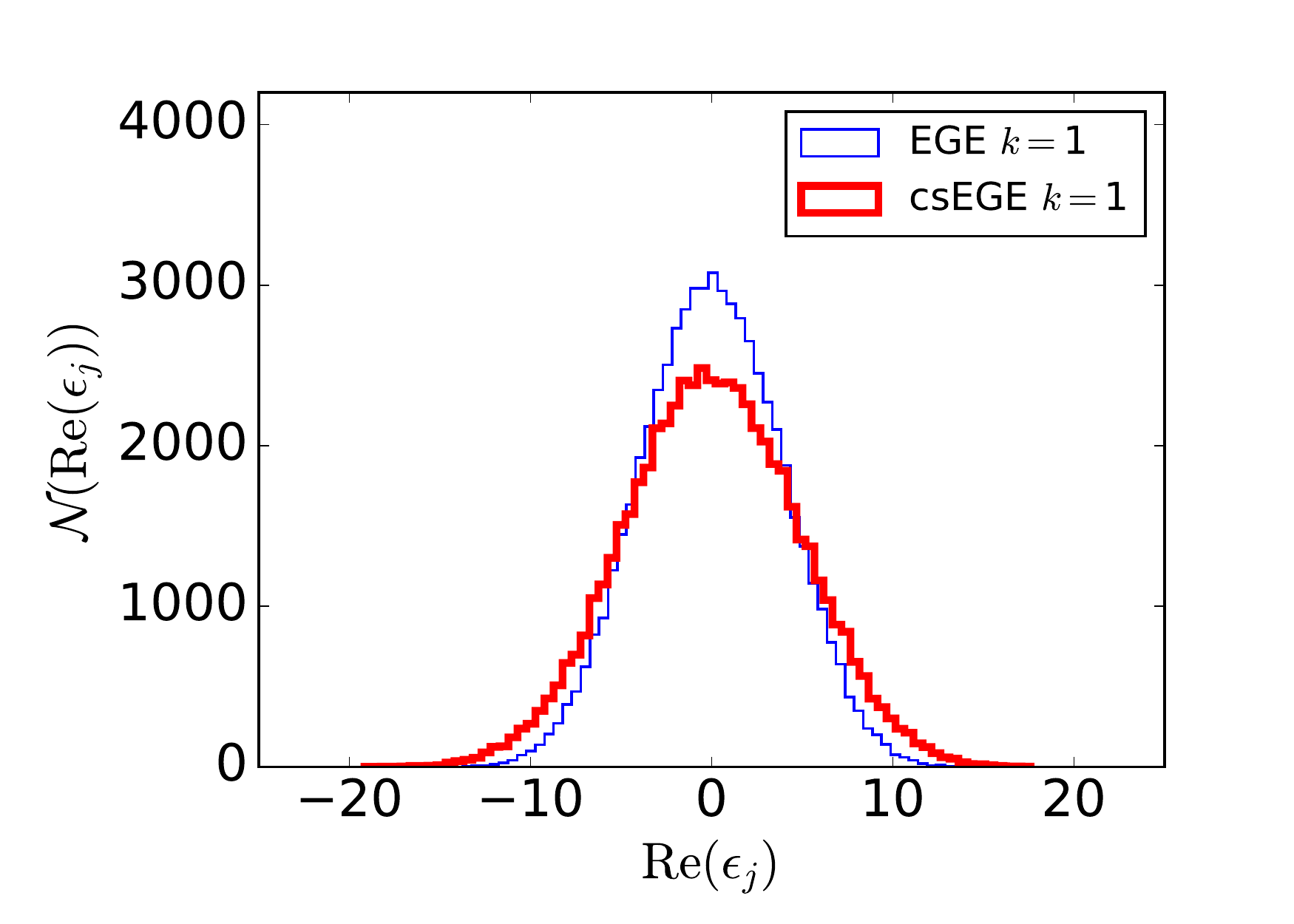}\\[-2mm]
  \includegraphics[scale=0.35]{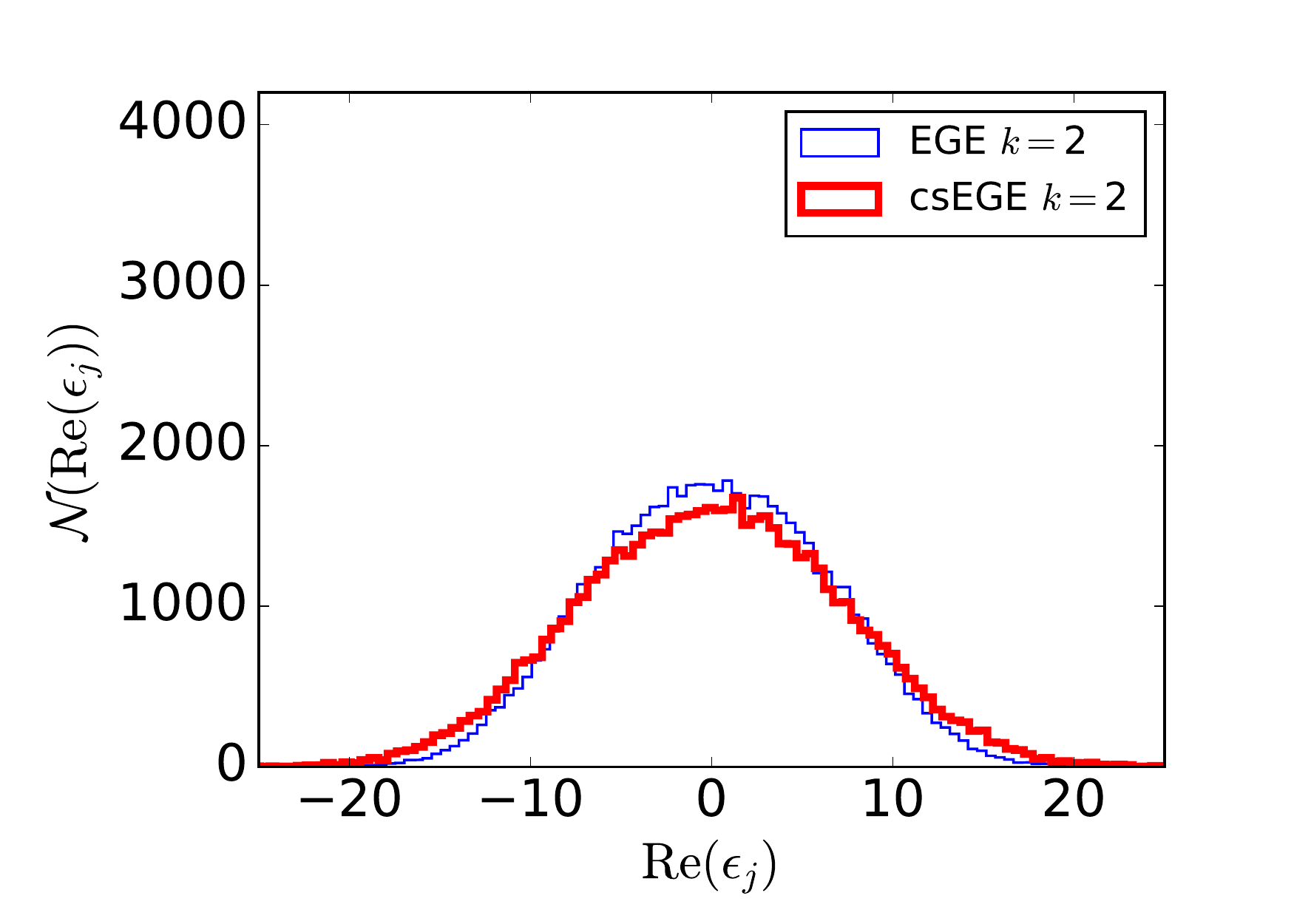}\\[-2mm]
  \includegraphics[scale=0.35]{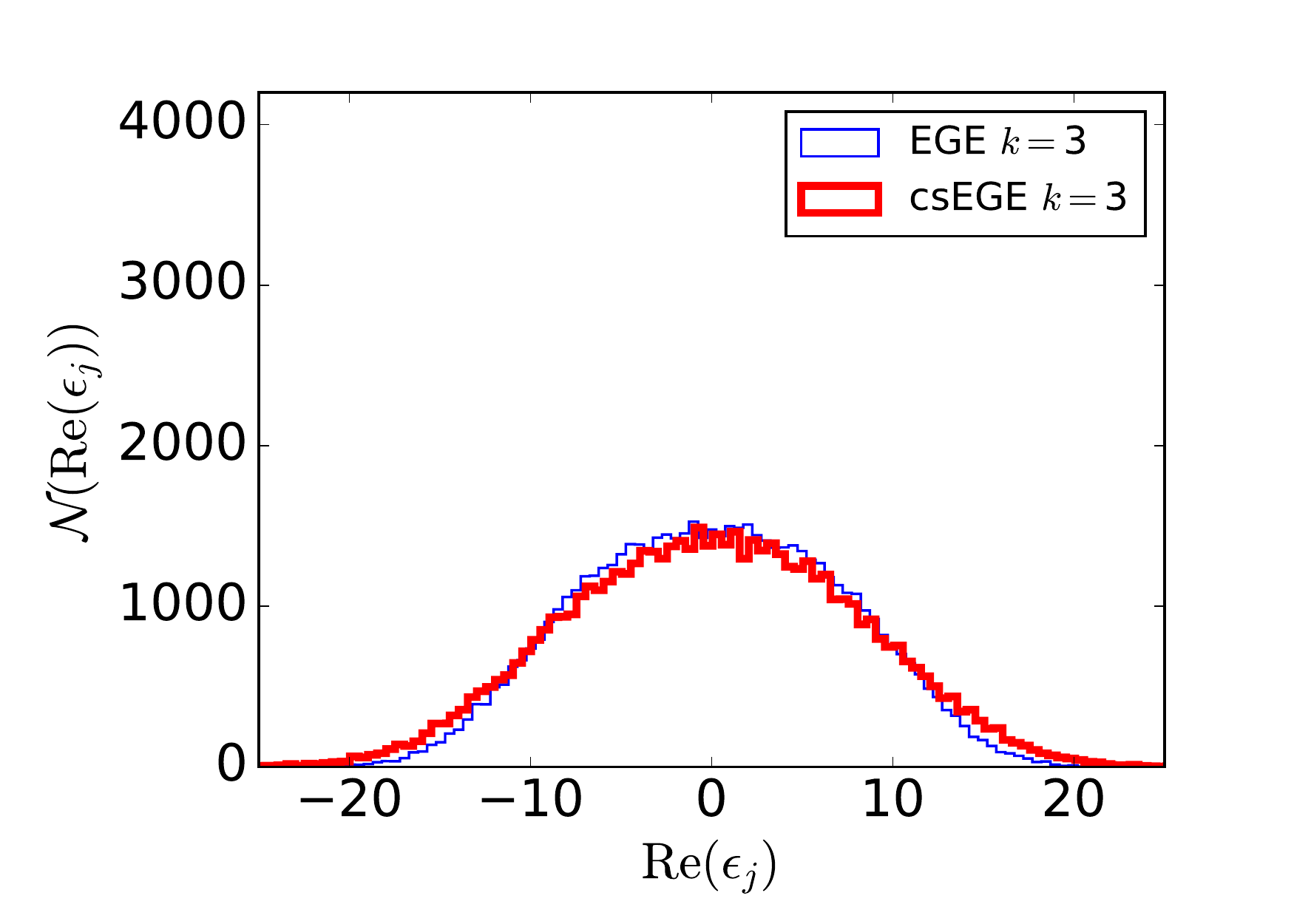}\\[-2mm]
  \includegraphics[scale=0.35]{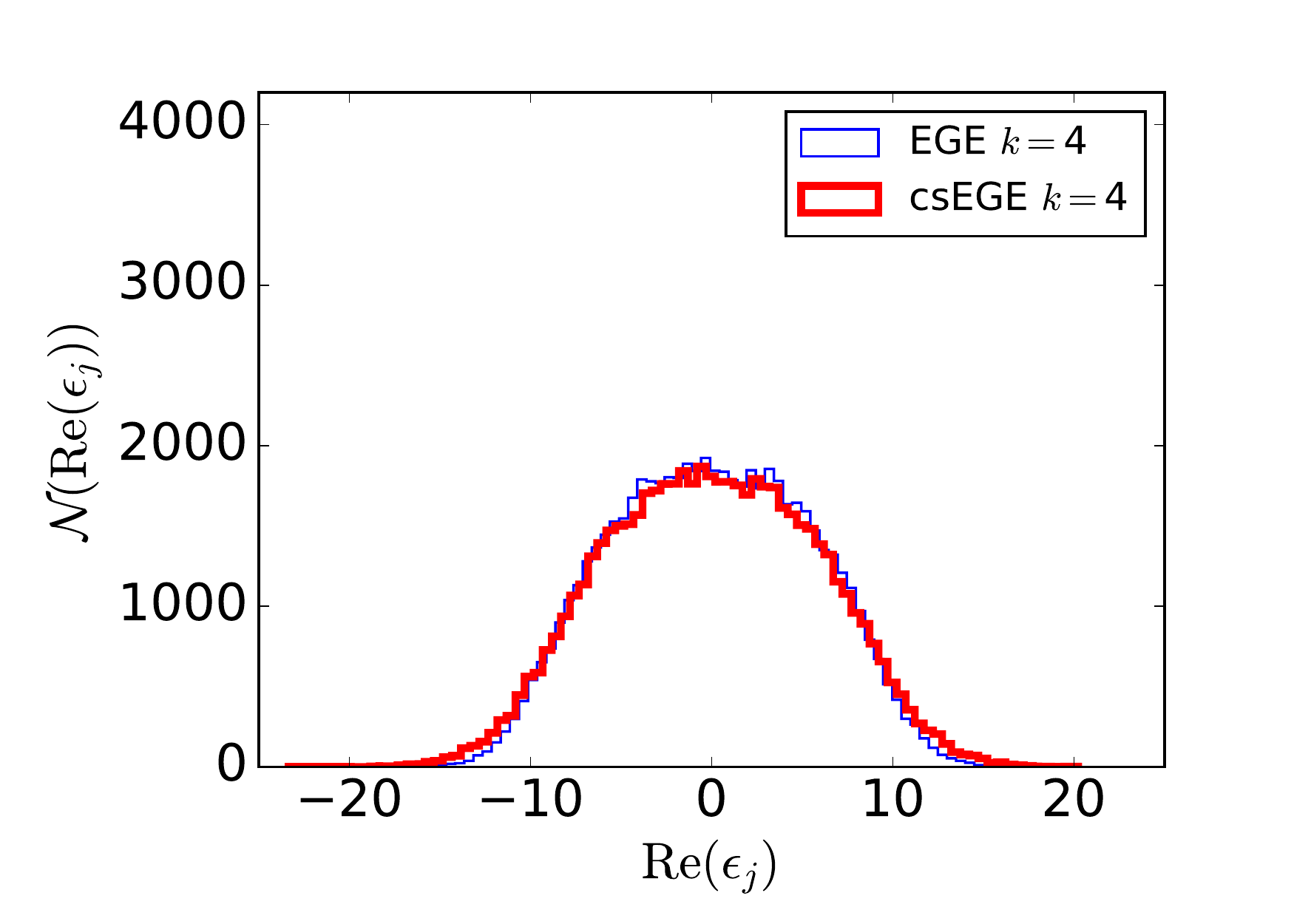}\\[-2mm]
  \includegraphics[scale=0.35]{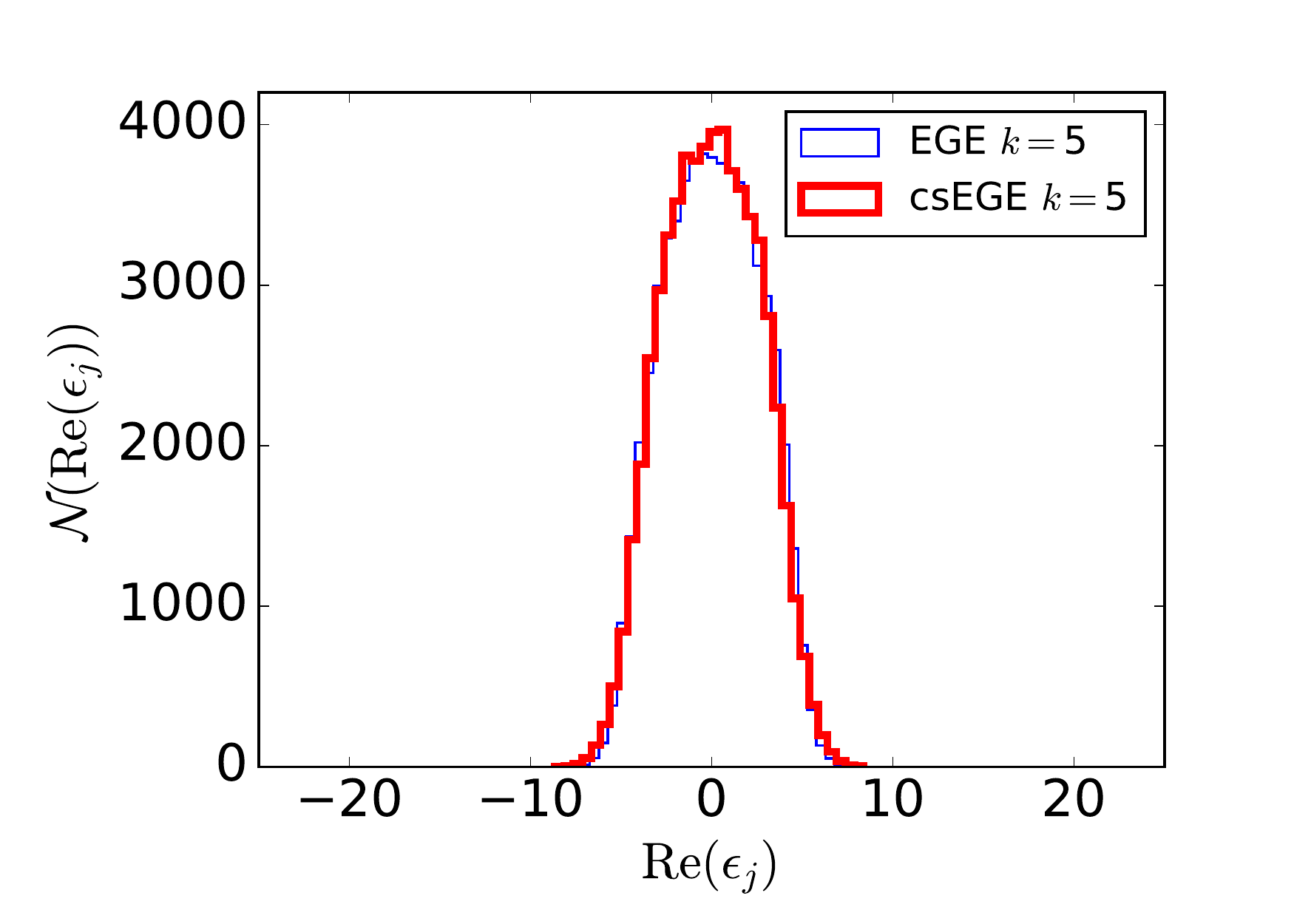}
  \caption{(color online) Histograms of $\Re{\eps_j}$ for $n=l-1=5$. The spectral span is maximal
    for $k \sim n/2$, which explains the maximum of the conductance bandwidth at this value, see
    \fig{3}. The csEGE (red-thick histogram) lead to a marginally wider spectral span but the differences decrease for
    increasing $k$.}
  \label{fig:7}
\end{figure}

The real part of the complex eigenvalues $\eps_j$ determine the position of the transmission
resonances; their distributions in terms of $k$, are shown in \fig{7}. We observe that
centrosymmetry has only weak effects; it increases marginally the spectral span. The differences
between the two cases are decreasing when $k$ increases. In both cases the spectral span is maximal
for $k \sim n/2$. These observations explain that the width of the conduction band is maximal for
$k\sim n/2$ and that it is marginally wider for the centrosymmetric case.

\begin{figure}
  \includegraphics[scale=0.38]{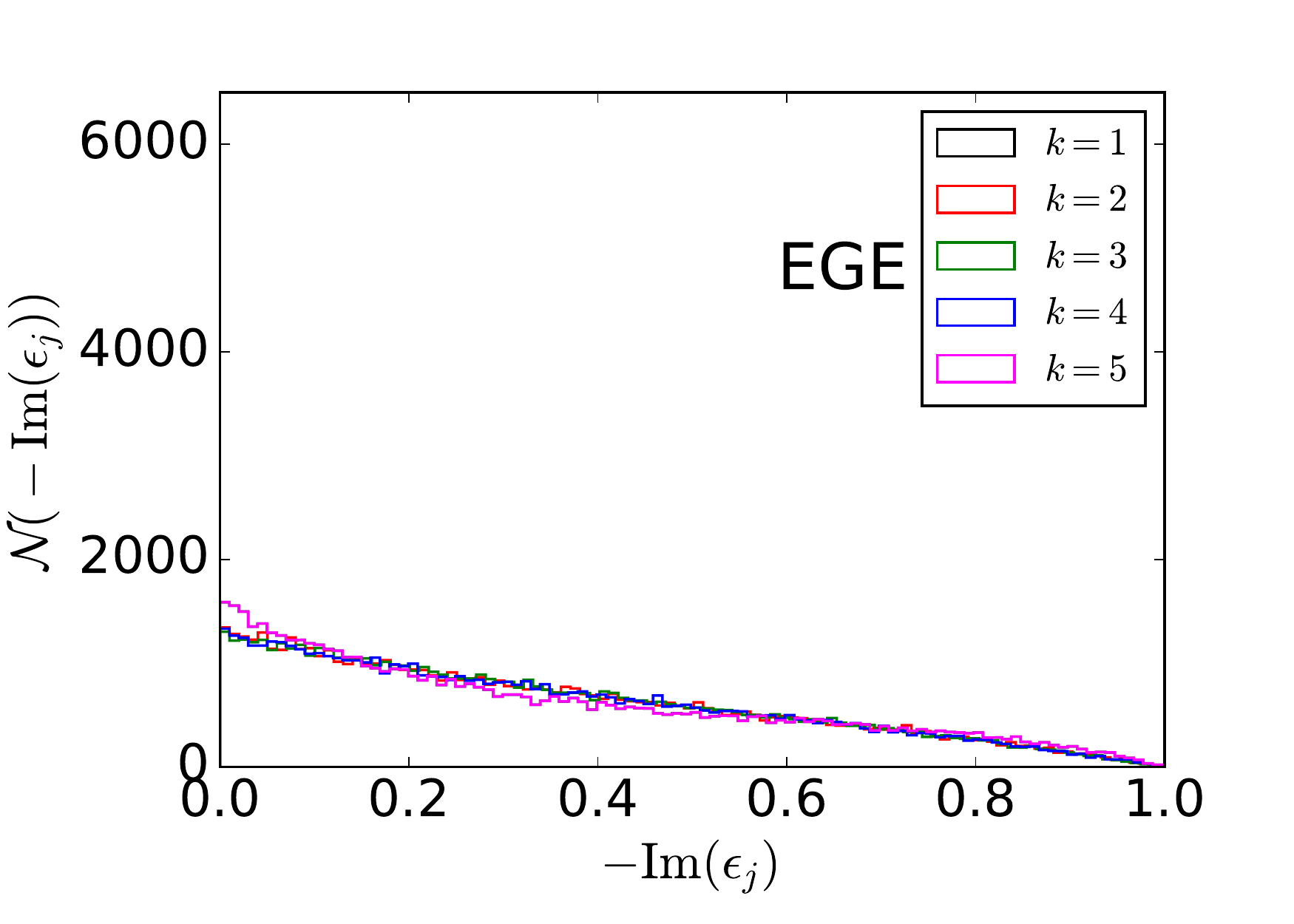}\\[-2mm]
  \includegraphics[scale=0.38]{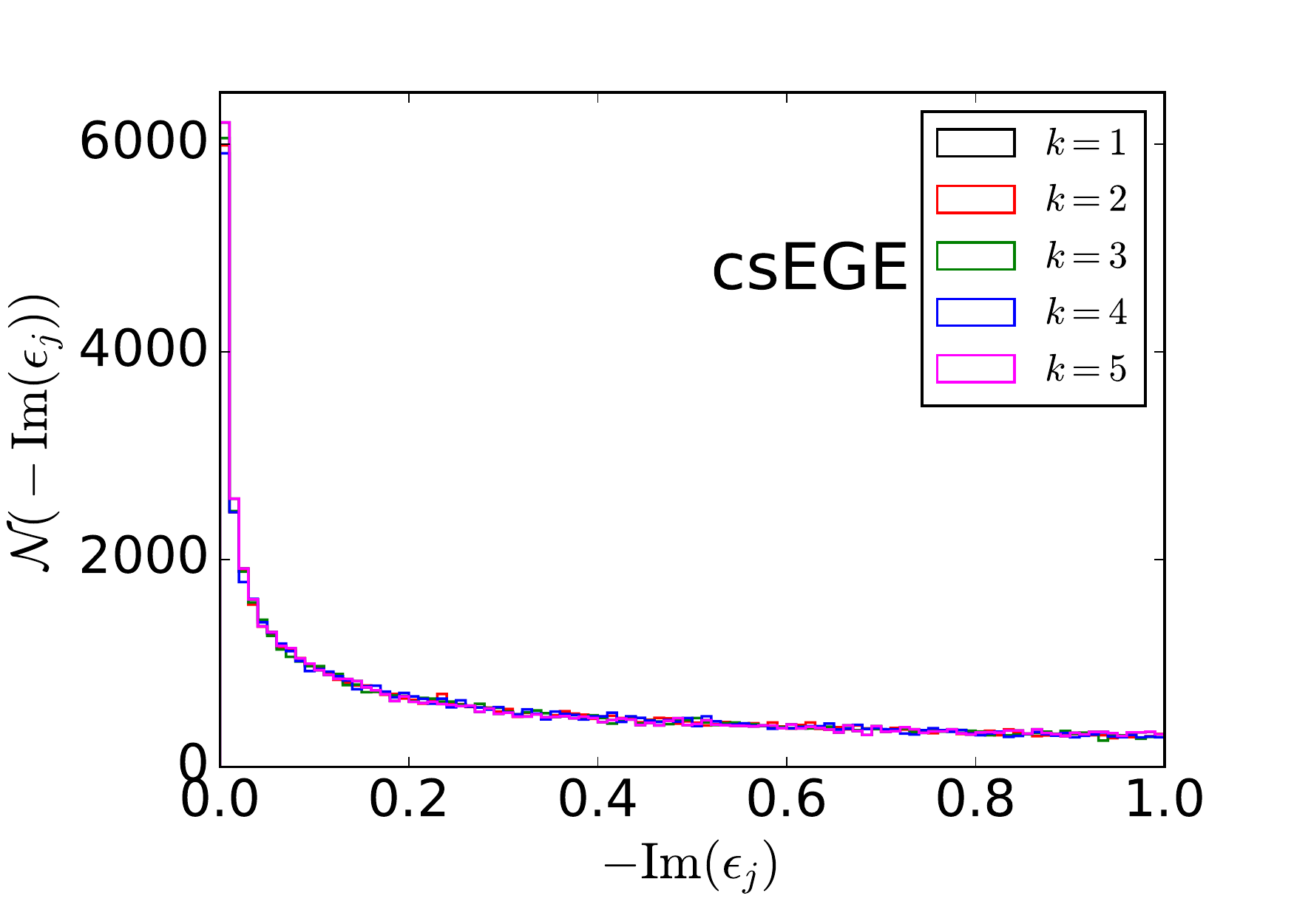}
  \caption{(color online) Distributions of $\Im{\eps_j}$ for $n=l-1=5$. The histograms are
    independent from the rank of interaction $k$, displaying a dependence on the presence or absence
    of centrosymmetry. Centrosymmetry enhances the number of eigenvalues with an imaginary part
    close to its minimum zero and its maximum $-\eta=-1$.}
  \label{fig:8}
\end{figure}

In turn, $\Im{\eps_j}$ is related to the width of the transmission peaks, see \fig{1}. The
distributions of $\Im{\eps_j}$ are presented in \fig{8}. As shown, their structure is essentially
independent of $k$. Comparing the EGE and csEGE cases we find that centrosymmetry amplifies the
occurrence of the extrema: the number of eigenvalues with $\Im{\eps_j}=0$ and $\Im{\eps_j}=-\eta=-1$
is larger when centrosymmetry is present. The former value corresponds to resonances of vanishing
width, while the latter is related to broad resonances. Then, the histograms for the EGE show that
the number of broad resonances vanishes linearly as $\Im{\eps_j}\to -\eta$, while for the csEGE
this limit attains a constant.

\begin{figure}
  \includegraphics[scale=0.38]{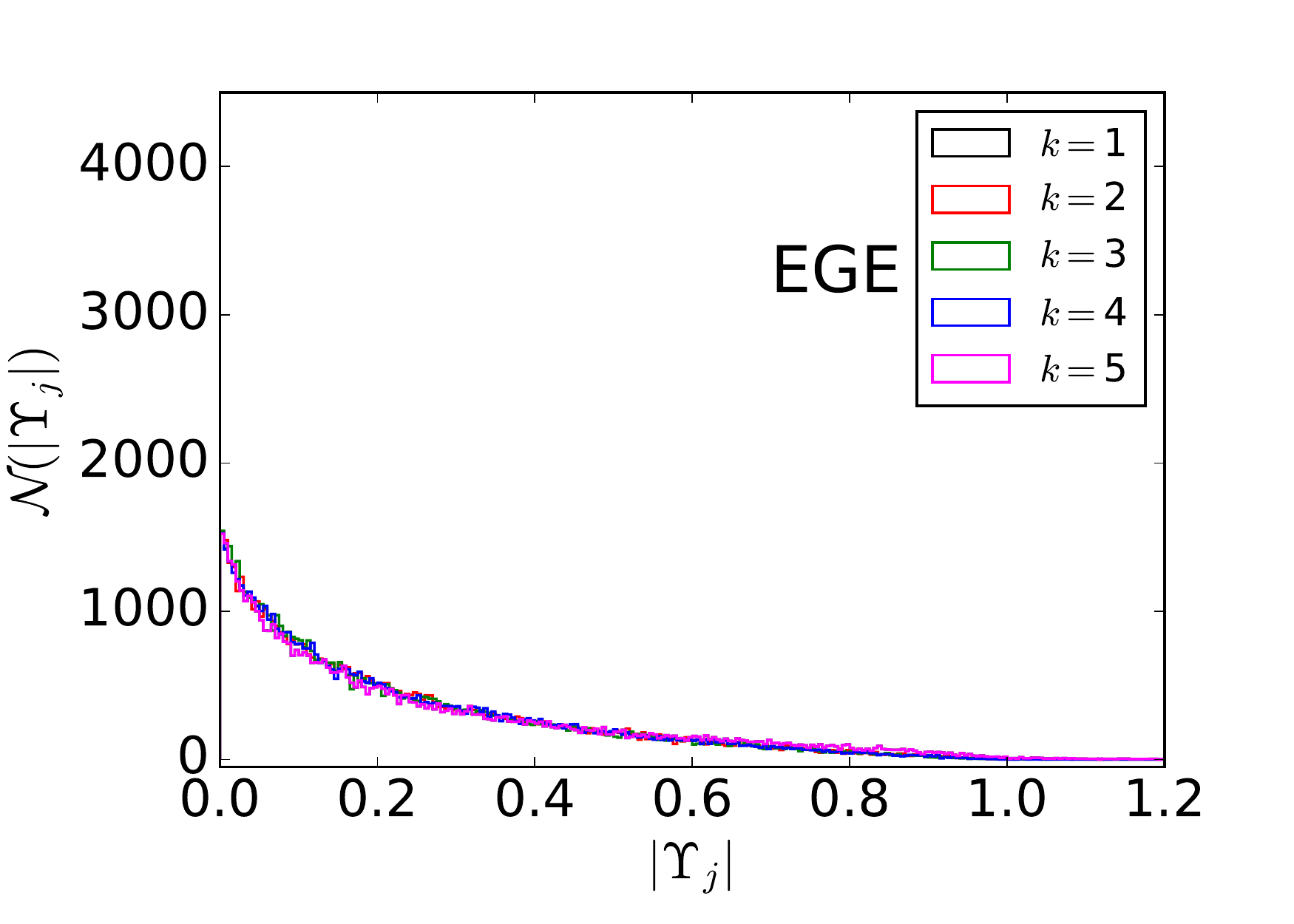}\\
  \includegraphics[scale=0.38]{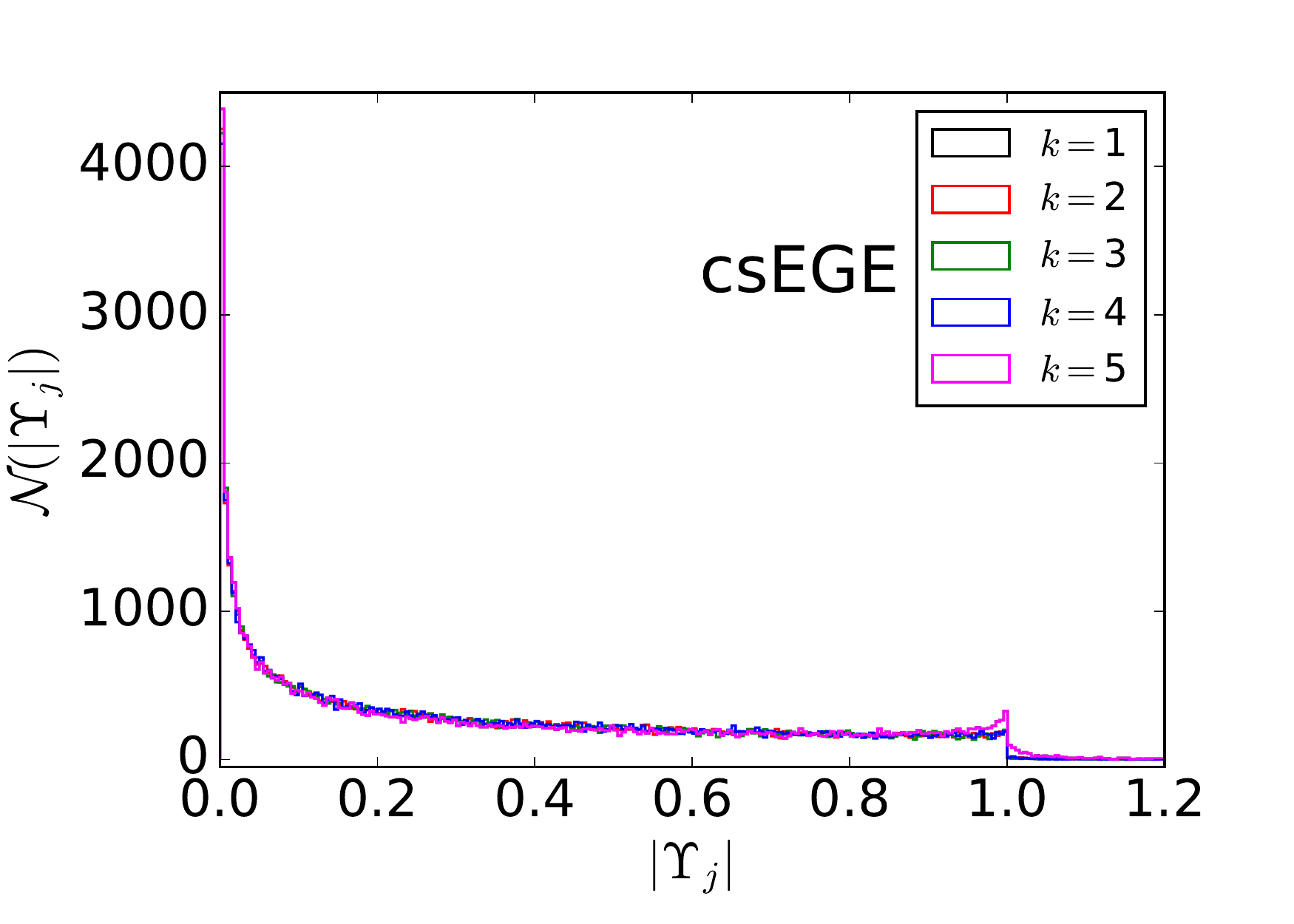}
  \caption{(color online) The histograms of $\Abs{\Upsilon_j}$ have similar properties as the
    histograms of $-\Im{\eps_j}$ in \fig{8}.}
  \label{fig:9}
\end{figure}

The corresponding distributions $\Upsilon_j$, see \eq{Ups}, is shown in \fig{9}. It displays similar
properties as the distributions of $\Im{\eps_j}$. That is, csEGE shows a larger frequency of events
displaying zero and the maximal values of $\Abs{\Upsilon_j}$ than the EGE, and the distributions
are essentially independent from $k$. This motivates us to investigate the correlations between
$\Abs{\Upsilon_j}$ and $\Im{\eps_j}$.

\begin{figure}
  \includegraphics[scale=0.38]{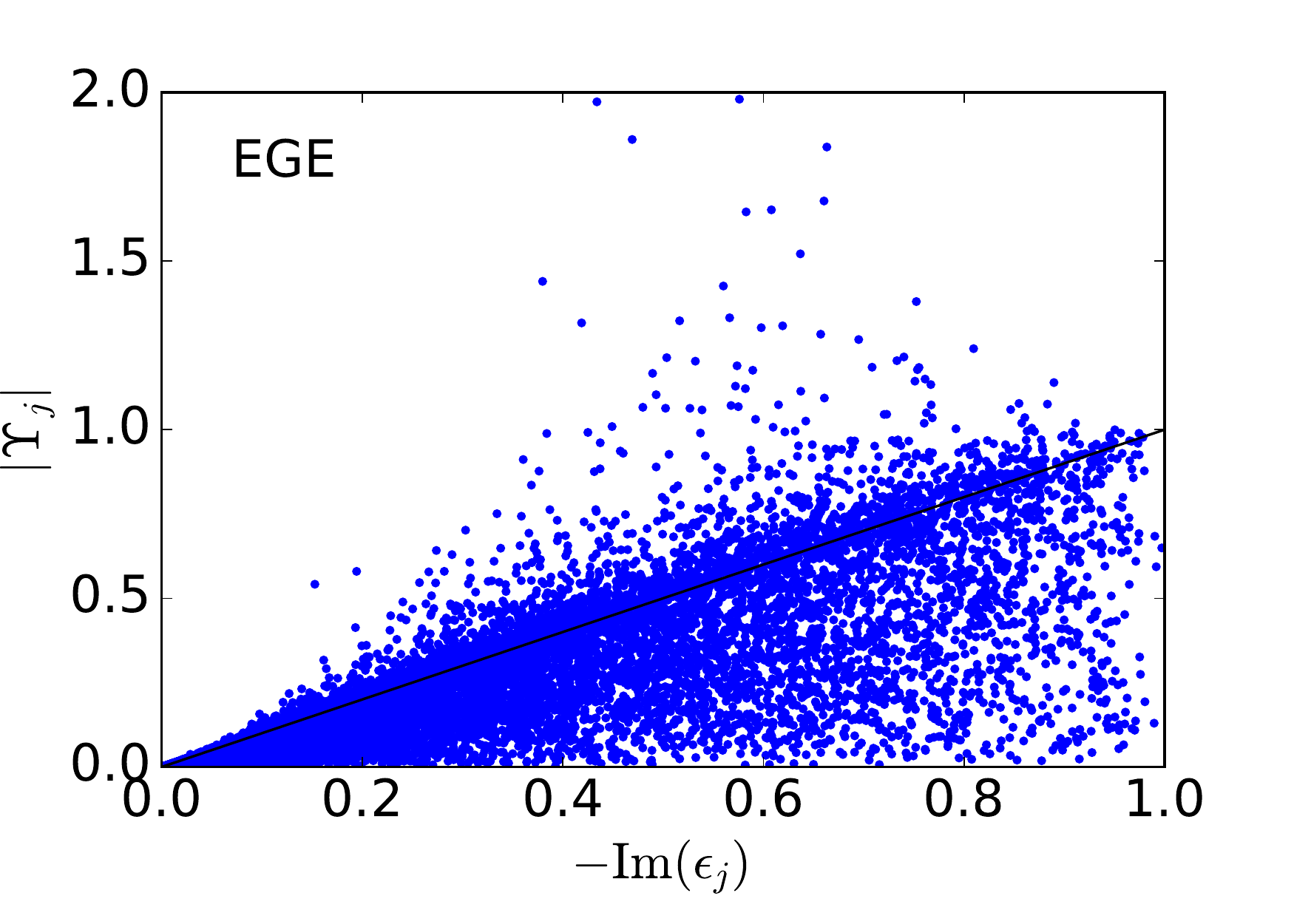}\\[-2mm]
  \includegraphics[scale=0.38]{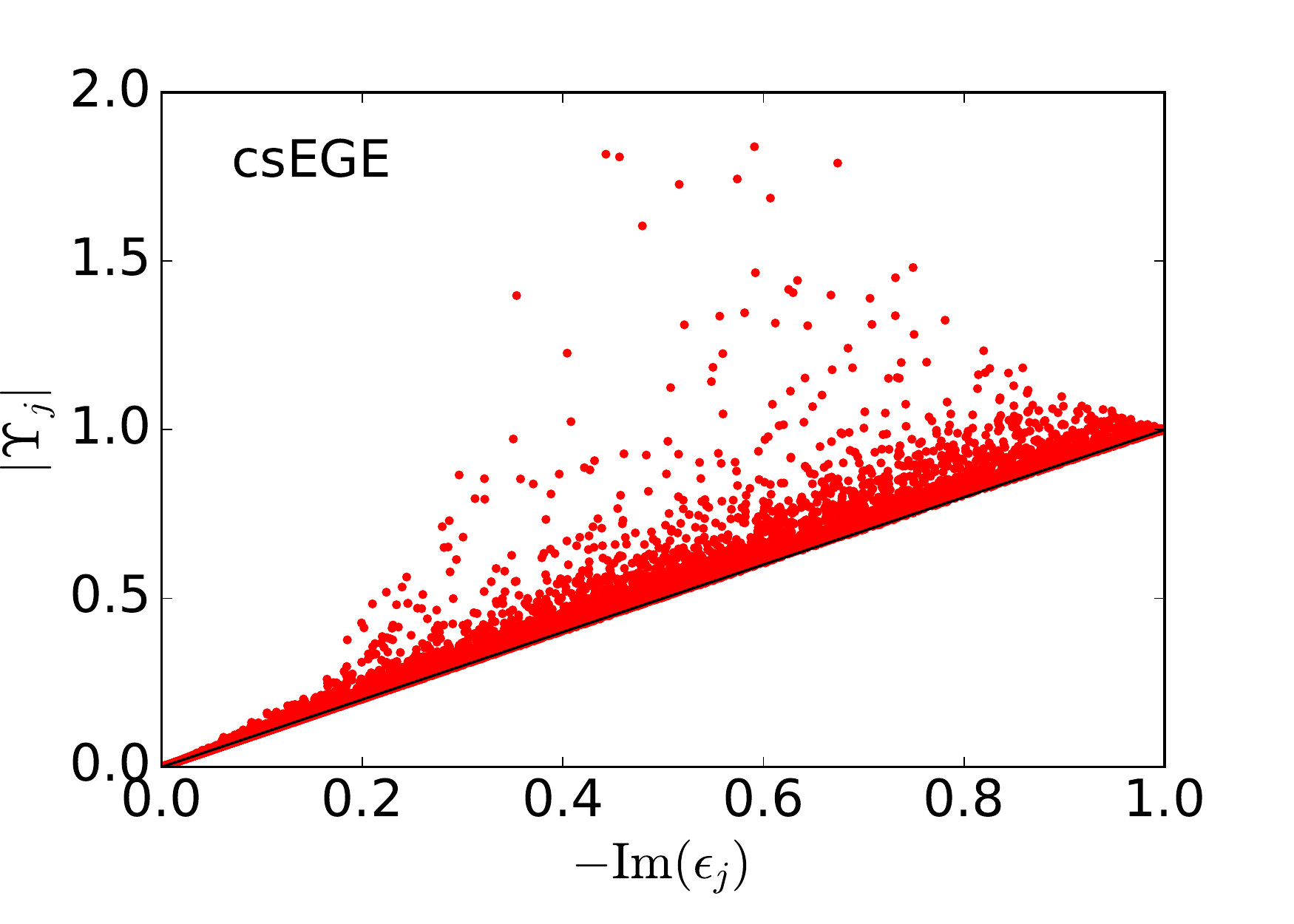}
  \caption{(color online) Distribution of $\Abs{\Upsilon_j}$ versus $-\Im{\eps_j}$ for ensemble of
    2000 realizations. Centrosymmetry imposes strong correlations. While in the case of arbitrary
    EGE the data points are located mainly in a triangle $\Abs{\Upsilon_j} \lesssim -\Im{\eps_j}$,
    in the case of csEGE the data points are pinned on the line $\Abs{\Upsilon_j} \sim -\Im{\eps_j}$
    or above it.}
  \label{fig:10}
\end{figure}

In \fig{10}, we plot $\Abs{\Upsilon_j}$ versus $\Im{\eps_j}$ illustrating strong correlations among
these quantities. While in the EGE case the data points are scattered in the triangular region
$\Abs{\Upsilon_j} \lesssim -\Im{\eps_j}$, in the case of the csEGE the data appear on the line
$\Abs{\Upsilon_j} \sim -\Im{\eps_j}$ or above it.

\begin{figure*}
  \includegraphics[scale=0.4]{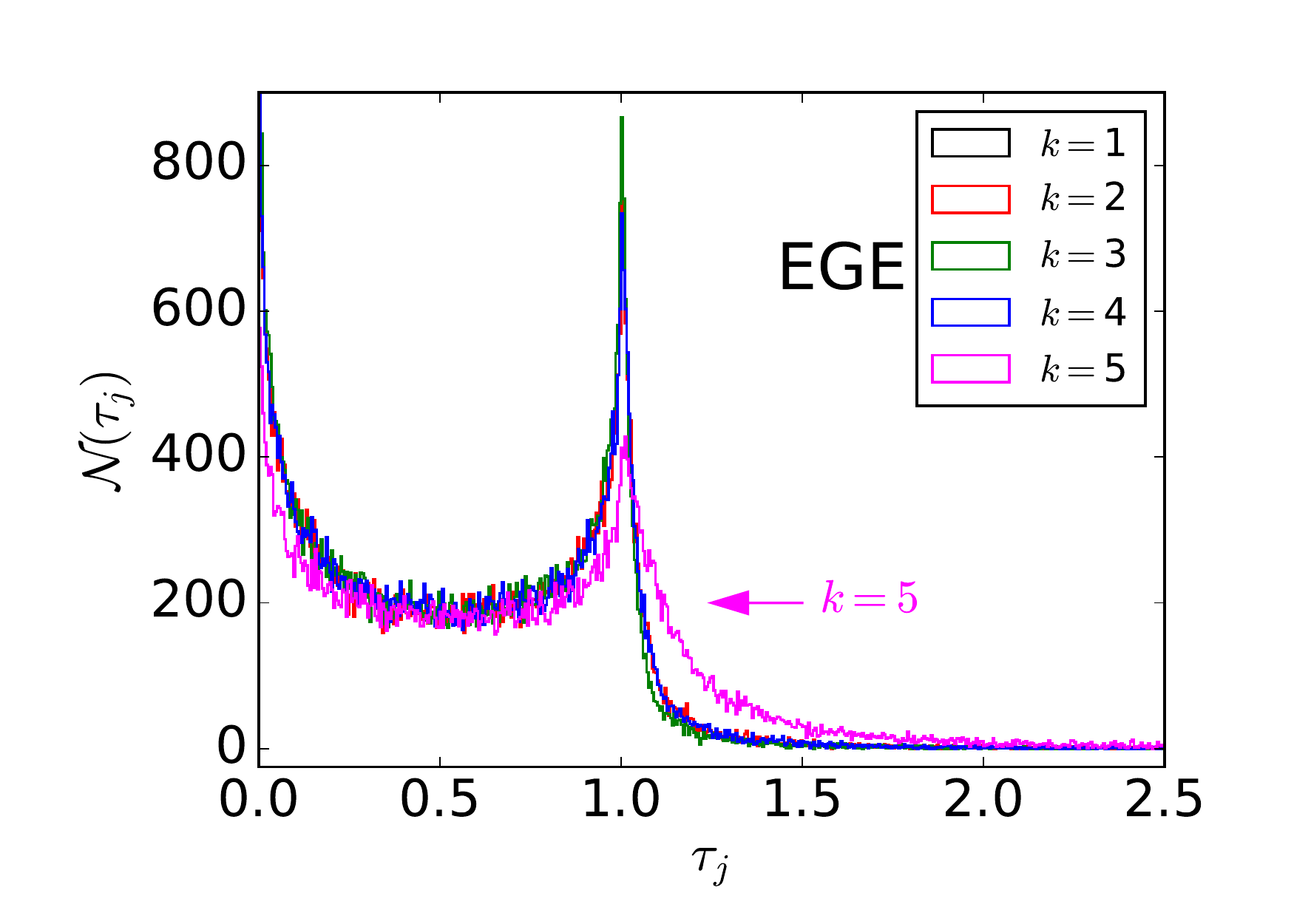}
  \includegraphics[scale=0.4]{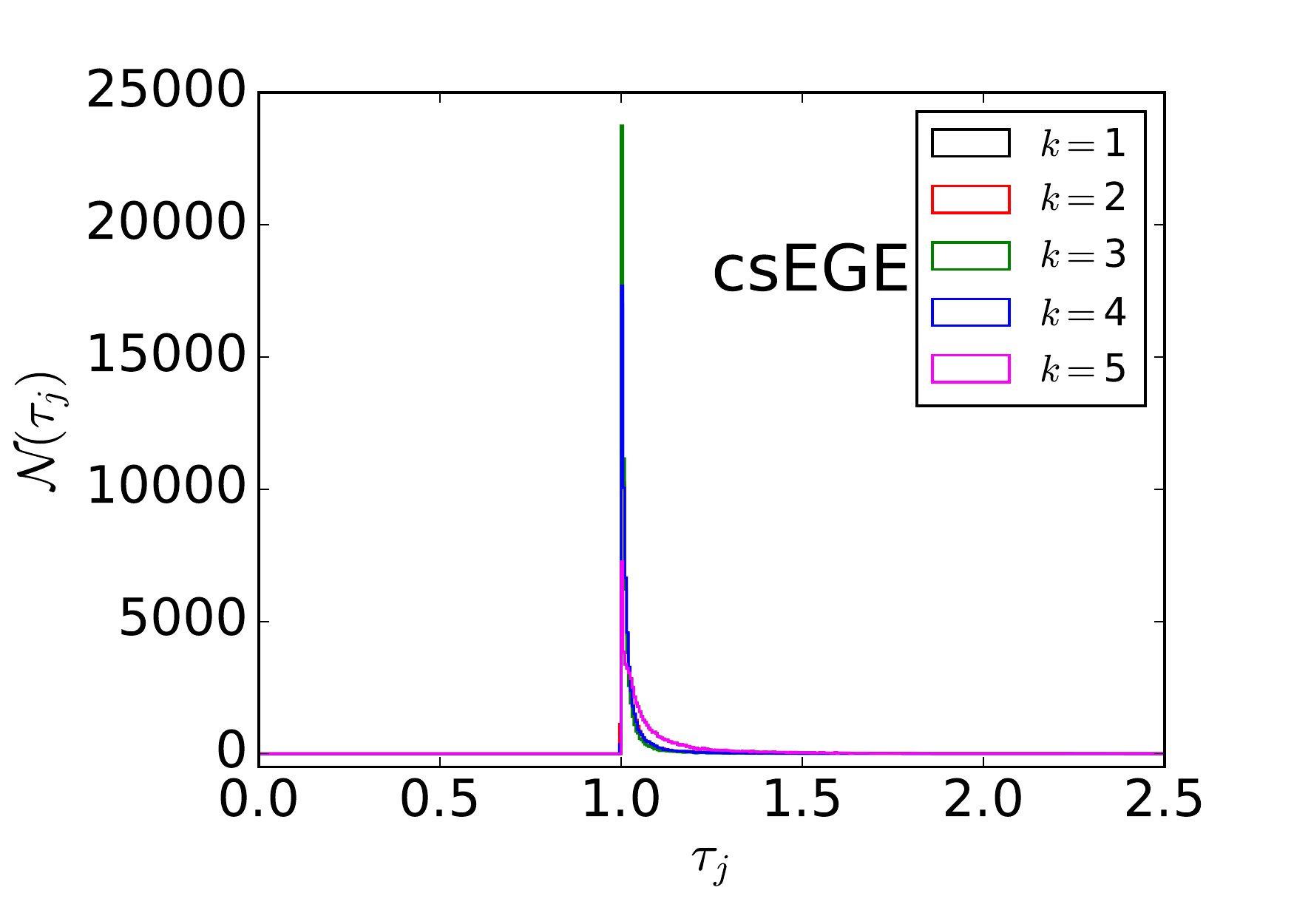}
  \caption{(color online) Histograms of the quotient $\tau_j\equiv \Abs{\Upsilon_j/\Im{\eps_j}}$.
    For EGE this quotient is distributed mainly between 0 and 1 with maxima at these points. In
    contrast, for csEGE the distribution of $\tau_j$ is pinned to 1 without any value less than
    1. This indicates the increase of resonances of perfect transmission due to centrosymmetry, see
    \fig{5}.}
\label{fig:11}
\end{figure*}

The histograms of the ratio $ \tau_j \equiv \Abs{\Upsilon_j/\Im{\eps_j}} $ in \fig{11} show these
strong correlations from another perspective. In view of \eq{Tdecomp}, the quantity $\tau_j$ yields
an estimate of the transmission by taking into account the main resonance only and neglecting all
interference effects, i.e. the other terms of the sum. For the EGE the distribution is mainly
located between 0 and 1, with peaks at these values, dominated especially by the $k\sim n/2$
case. In turn, for the csEGE the values of $\tau_j$ are peaked strongly at 1 with a decaying tail
beyond 1 but without any $\tau_j$ smaller than 1. Note also the difference in the vertical
scales. The $\tau_j$ may attain values larger than 1 because the phases are neglected, which cause
the transmission to be equal or less than 1. These two histograms close our statistical analysis to
understand how centrosymmetry enhances transport. They confirm that centrosymmetry enhances the
extrema and induces strong correlations which generate numerous transmission resonances of perfect
transport ($T=1$), see \fig{5}.

\section{Conclusions \& Outlook}
\label{sec:ConclusionsOutlook}

In this paper, we have analyzed the stationary transport properties of fermions through small
disordered interacting networks, which are modeled by embedded Gaussian ensemble of random
matrices. We have addressed the influence of centrosymmetry and have shown that the transport is
enhanced significantly if centrosymmetry is present. This applies for the transmission
$\braket{T(E)}$, which is a function of the energy of the excitation, as well as for the averaged
total current $\braket{I}$ through the system, see \fig{3} and \fig{4}, respectively.

We have shown that centrosymmetry induces many transmission resonances of perfect transport
(i.e. $T=1$) which enhances the transport in the overall conduction band, see \fig{1} and
\fig{5}. We have also seen that ---independent from the fact whether the system is centrosymmetric
or not--- the width of the conduction band is maximal for $k \sim n/2$ and increases with $n$ until
the system is almost filled. In contrast, for $k=1$ and $k=n$ the conduction band width is
minimal. In the best case, which appears when the system is almost filled $n=l-1$ and the rank of
interaction is $k\sim n/2$, centrosymmetry enhances the averaged total current by $75\%$ and its
mode increases by a factor of two. For larger systems, see the Appendix, the best cases appear for
values of $n$ close but less than $l-1$; the improvement of the transport by centrosymmetry is even
stronger. Moreover, we observe that the distribution of the total current for the csEGE has a very
large dominating peak for $n=l-1$, close to the highest observed currents. Our results are therefore
in prefect agreement with our previous work \cite{2015ADP-OrtegaMananBenet}.

Using the spectral decomposition of the Green's function, we have shown that centrosymmetry enhances
the extrema, see \fig{8} and \fig{9}. The number of resonances with minimal (0) and maximal
(1) width (proportional to $\Im(\eps_j)$) and weight ($\propto \Abs{\Upsilon_j}$)
increases. Centrosymmetry also induces strong correlations between the width and the weight of the
resonances, see \fig{10} and \fig{11}. This suppresses destructive interference effects in the
system and thus, causes backscattering-free transmission resonances which enhance the overall
transport. We interpret these results as a manifestation that centrosymmetry is an important
property for the design of quantum networks with efficient transport characteristics.

Comparing with \cite{2013PRL-walschaers}, we find that a doublet-structure in the Hamiltonian is not
required to improve the transport in the system. Centrosymmetry, which is imposed at the one
particle level and taken to the $k$ and $n$ particle space, and the $k$-body interaction, are enough
to enhance significantly the transport characteristics of the system.

Some degree of decoherence will be inevitably present in biomolecules at room temperature and, to
some extent, also in quantum devices. Therefore, we are currently extending the model to study the
effects of decoherence on the transport efficiency using the statistical approach which we have
developed recently \cite{Stegmann2012, Stegmann2014, StegmannPhD2014}.

\begin{acknowledgments}
  We are very grateful to T. Gorin and  M. Vyas for useful discussions and helpful
  remarks. T.S. acknowledges a postdoctoral fellowship from DGAPA-UNAM. Financial support from
  CONACyT research grant 219993 and PAPIIT-DGAPA-UNAM research grant IG100616 is acknowledged.
\end{acknowledgments}

\appendix*
\section{Systems with $l=8$ and $l=10$ single-particle states}
\label{sec:appendix}

In this Appendix we show, equivalently to \fig{3} and \fig{4}, the ensemble averaged transmission
and the frequency histogram of the total current for systems consisting of $l=8$ and $l=10$
single-particle states. Note that although we have added only up to 4 single-particle states to the
system studied in the main text, the dimension of the Hilbert space $ \binom{l}{n} $ is up to 10
times larger. Note also that the spectral span, the width of the conduction band and hence, the
total current increase with $l$, see \cite{2001AnnPhys-BRW}. We observe qualitatively the same
properties as for the smaller system with $l=6$, except that maximal values of the total current are
observed also for fillings close but less than $l-1$. The mode of the distribution of the total
current for the csEGE is strongly peaked for $n=l-1$, close to the highest observed currents. The
statistics of the spectral decomposition is the same as for the system discussed in the main text.

\begin{turnpage}
  \begin{figure*}
    \includegraphics[scale=0.35]{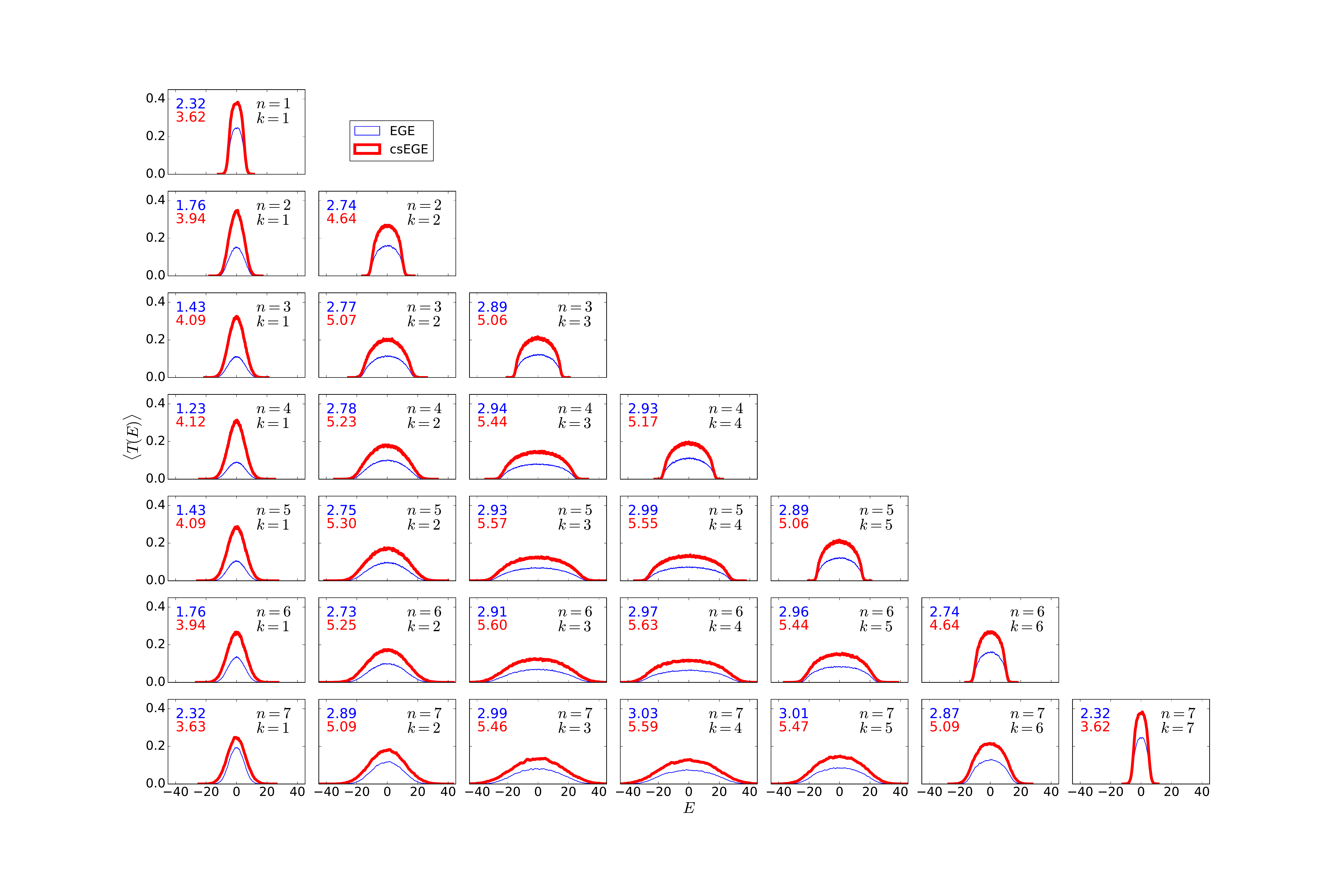}
    \caption{(color online) Ensemble averaged transmission $\braket{T(E)}$ for a system with $l=8$
      single-particle states. The arrangement of the figures is the same as in \fig{3}.}
    \label{fig:12}
  \end{figure*}

  \begin{figure*}
    \includegraphics[scale=0.35]{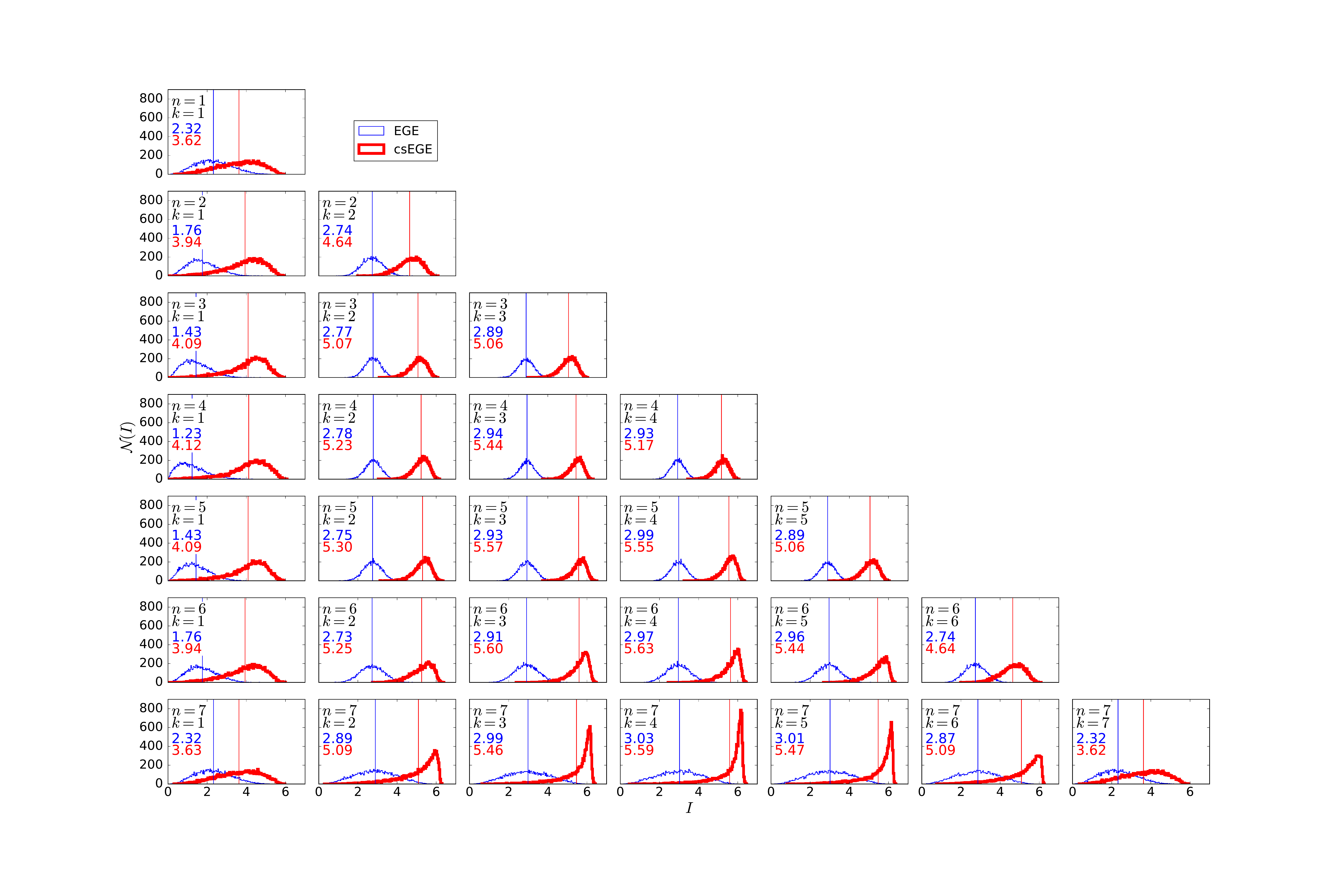}
    \caption{(color online) Frequency histogram of the current for a system with $l=8$. The
      arrangement of the figures is the same as in \fig{4}.}
    \label{fig:13}
  \end{figure*}

  \begin{figure*}
    \includegraphics[scale=0.23]{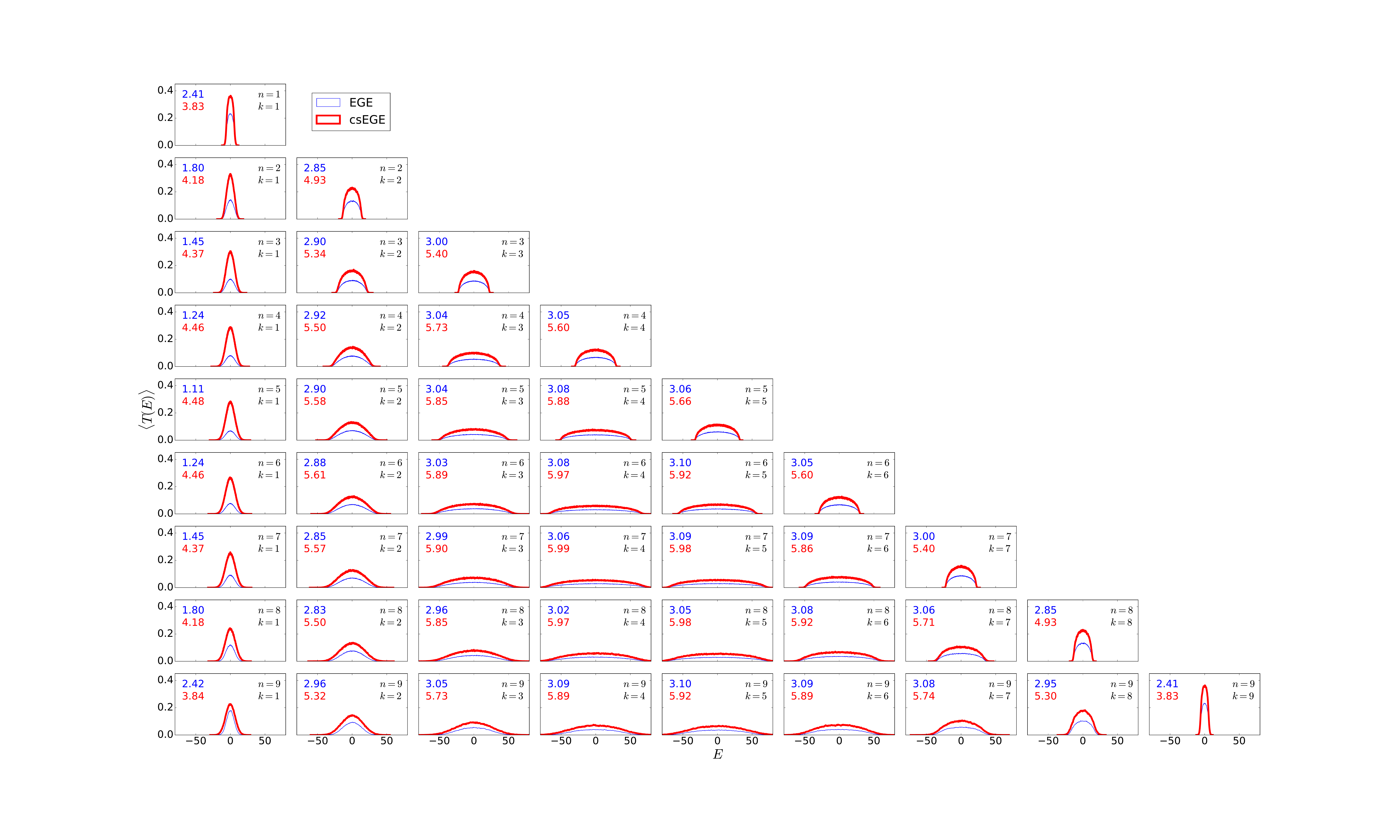}
    \caption{(color online) Ensemble averaged transmission $\braket{T(E)}$ for a system with
      $l=10$. The arrangement of the figures is the same as in \fig{3}.}
    \label{fig:14}
  \end{figure*}

  \begin{figure*}
    \includegraphics[scale=0.23]{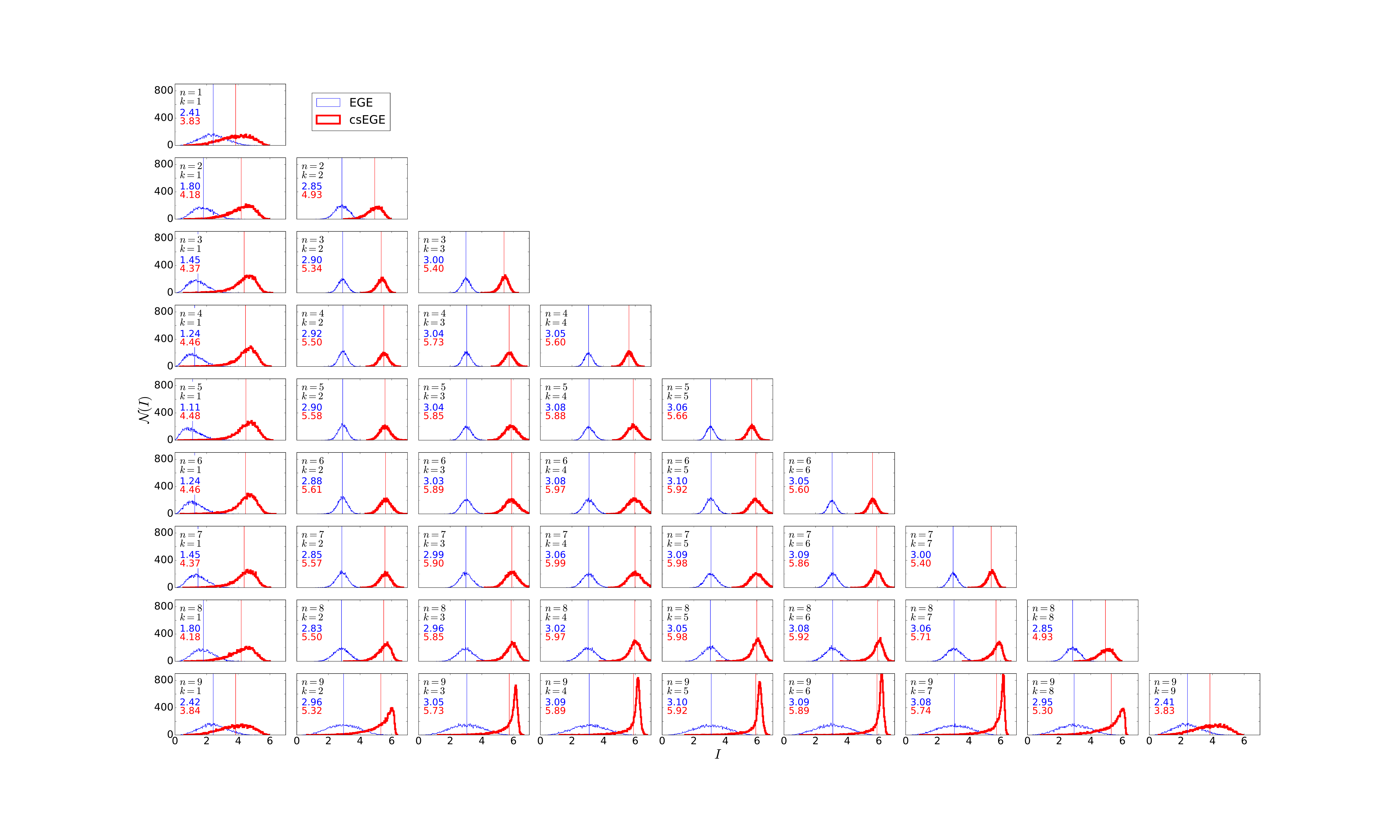}
    \caption{(color online) Frequency histogram of the current for a system with $l=10$. The
      arrangement of the figures is the same as in \fig{4}.}
    \label{fig:15}
  \end{figure*}
\end{turnpage}

\bibliography{bibliografia.bib}

\end{document}